\documentclass[pre,showkeys,preprintnumbers,amsmath,amssymb,superscriptaddress,twocolumn]{revtex4}
\pdfoutput=1
\usepackage{graphicx}
\usepackage{amstext}
\usepackage{makeidx}
\usepackage{amssymb,amsfonts}
\usepackage{verbatim}
\usepackage{epstopdf}
\usepackage{hyperref} 
\usepackage{color}
\usepackage{bm}
\usepackage{comment}

\def\P{{\mathbb P}}
\def\x{\mathbf{x}}

\def\q{\mathbf{q}}
\def\s{\mathbf{s}}
\def\R{{\mathbb R}}

\def\var{\textrm{var}}

\def\erf{\mathrm{erf}}

\def\K{{\mathcal K}}

\def\pa{\partial\Omega}

\def\T{{\mathcal T}}

\def\SIgeneral{1~}
\def\SISS{2~}
\def\SIlimit{3~}
\def\SIunbounded{4}
\def\SIsubordination{5}

\begin{document}

\title{Diffusion-limited reactions in dynamic heterogeneous media}

\author{Yann Lanoisel\'ee}
\affiliation{Laboratoire de Physique de la Mati\`{e}re Condens\'{e}e (UMR 7643), \\ 
CNRS -- Ecole Polytechnique, University Paris-Saclay, 91128 Palaiseau, France}

\author{Nicolas Moutal}
\affiliation{Laboratoire de Physique de la Mati\`{e}re Condens\'{e}e (UMR 7643), \\ 
CNRS -- Ecole Polytechnique, University Paris-Saclay, 91128 Palaiseau, France}

\author{Denis S. Grebenkov}
 \email{denis.grebenkov@polytechnique.edu}
\affiliation{Laboratoire de Physique de la Mati\`{e}re Condens\'{e}e (UMR 7643), \\
CNRS -- Ecole Polytechnique, University Paris-Saclay, 91128 Palaiseau, France}

\date{\today}

\begin{abstract}
Most biochemical reactions in living cells rely on diffusive search
for target molecules or regions in a heterogeneous overcrowded
cytoplasmic medium.  Rapid re-arrangements of the medium constantly
change the effective diffusivity felt locally by a diffusing particle
and thus impact the distribution of the first-passage time to a
reaction event. Here, we investigate the effect of these dynamic
spatio-temporal heterogeneities onto diffusion-limited reactions.  We
describe a general mathematical framework to translate many results
for ordinary homogeneous Brownian motion to heterogeneous diffusion.
In particular, we derive the probability density of the first-passage
time to a reaction event and show how the dynamic disorder broadens
the distribution and increases the likelihood of both short and long
trajectories to reactive targets.  While the disorder slows down
reaction kinetics on average, its dynamic character is beneficial for
a faster search and realization of an individual reaction event
triggered by a single molecule.
\end{abstract}


\maketitle

Diffusion is the central transport mechanism in living cells and, more
generally, in biological systems.  Molecular overcrowding,
cytoskeleton polymer networks and other structural complexities of the
intracellular medium lead to various anomalous features such as
nonlinear scaling of the mean square displacement (MSD), weak
ergodicity breaking, non-Gaussian distribution of increments, or
divergent mean first passage times (FPT) to reactive targets
\cite{Weiss2004,Loverdo2008,Wang2009,Benichou2010,Wang2012,Bertseva2012,Barkai2012,Manzo2015,He2016,Sadegh2017}.  
These features are often captured in theoretical models via long-range
correlations (e.g., fractional Brownian motion or generalized Langevin
equation), long-time caging (continuous time random walks), or
hierarchical structure (diffusion on fractals)
\cite{Bouchaud1990,Metzler2000,Havlin2002,Sokolov2012,Hofling2013,Bray2013,Metzler2014,Benichou2014}.
While the impact of heterogeneity of the medium
\cite{Torquato,Ghosh2015,Ghosh2016} and of reactive sites
\cite{Kayser1983,Weiss1986} onto diffusion and the macroscopic reaction
rate was investigated, the diffusivity of a particle was usually
considered as constant.  However, the structural organization of
living cells and other complex systems such as colloids, actin gels,
granular materials, and porous media suggests that the diffusivity
can vary both in space and time.

Several recent studies were devoted to such heterogeneous diffusion
models.  At the macroscopic level, the dynamics and the reaction
kinetics can still be described by the Fokker-Planck equation but time
and particularly space dependence of diffusivity prevents from getting
exact explicit solutions, except for some very elementary cases.
Moreover, in structurally disordered media, variations of diffusivity
are random, and the need for averaging over random realizations of the
disorder makes theoretical analysis particularly challenging.  Two
typical situations are often investigated.  If the disordered medium
is immobile (or changes over time scales much longer than that of the
diffusion process), the space-dependent diffusivity is considered as a
static field, in which diffusion takes place.  Whether the diffusivity
field is deterministic or random, its spatial profile can
significantly impact the diffusive dynamics and, in particular, the
distribution of the first-passage time to a reaction event
\cite{Fa2005,Cherstvy2013,Vaccario2015,Godec2016a,Tupikina2018}.
Note that the situation with a random static diffusivity is referred
to as ``quenched disorder'' and enters into a family of models known
as ``random walks in random environments''
\cite{Hughes,Murthy1989,Bouchaud1990b,Valle1991,Dean2014,Luo2015}.

\begin{figure}
\begin{center}
\includegraphics[width=85mm]{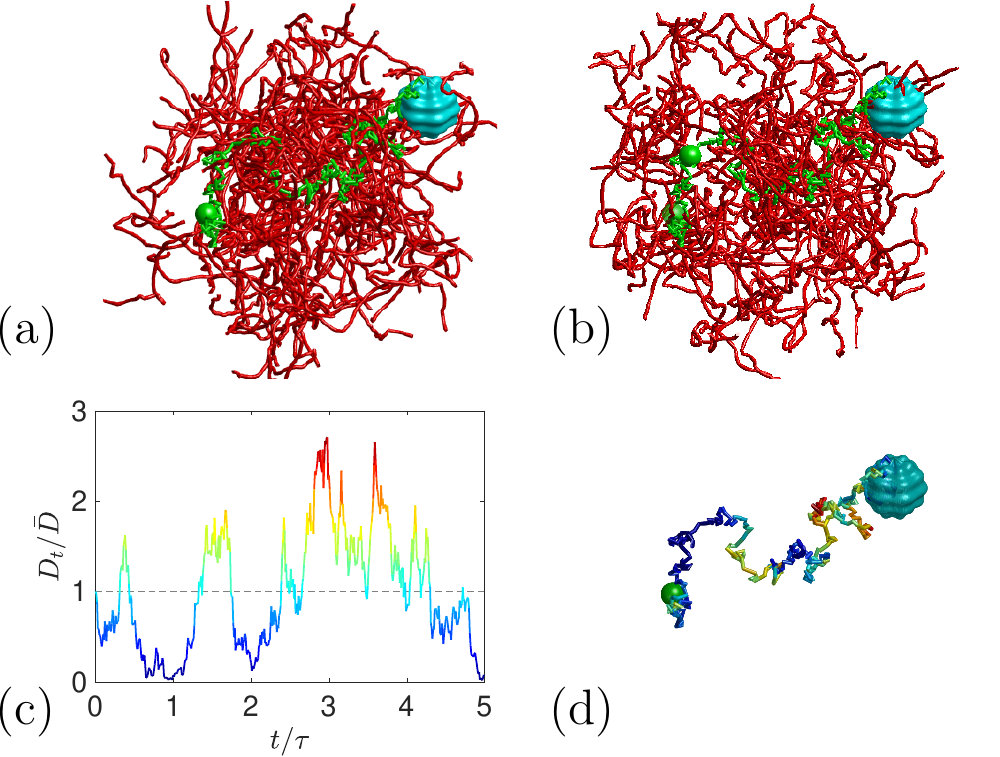}
\end{center}
\caption{
{\bf Schematic illustration of the annealed disorder model.}  A
dynamic heterogeneous medium is formed by a re-arranging polymer
solution (red thin tubes mimicking, e.g., actin filaments): {\bf
(a,b)} two snapshots of a particle (small green ball) diffusing toward
a reactive site (light blue bumpy object mimicking, e.g., a protein);
a random path (in green) of this particle is added to guide eyes;
along the path, the particle interacts with the local environment
re-arranging on a time scale $\tau$ and thus experiences variable
effective diffusivities; {\bf (c)} the environment-induced
time-dependent diffusivity $D_t$ is modeled by the Feller process
(\ref{eq:Feller}); {\bf (d)} once the re-arranging environment is
taken into account via $D_t$, one deals with the random path from the
initial position of the particle (green ball) to the target; the path
is explored with a time-dependent ``speed'' $D_t$, encoded by color as
in panel {\bf (c)}.}
\label{fig:actin}
\end{figure}

In turn, when the medium changes faster than the diffusion time scale,
a particle returning to a previously visited point would probe a
different local environment that can be modeled by a new realization
of random diffusivity at that point.  For instance, when a large
protein or a vesicle diffuses inside a living cell, other
macromolecules, actin filaments and microtubules can move
substantially on comparable time scales, changing the local
environment \cite{Sadegh2017,McGuffee2010,Yu2016,Samanta2016} (see
Fig. \ref{fig:actin}).  It is thus natural to consider the diffusivity
as a stochastic time-dependent process, $D_t$, referred to as
``annealed disorder''.  The concept of ``diffusing diffusivity'' was
put forward by Chubynsky and Slater \cite{Chubynsky2014} and then was
further developed by Jain and Sebastian \cite{Jain2016,Jain2017} and
Chechkin {\it et al.} \cite{Chechkin2017} (note that the impact of a
stochastic volatility onto the distribution of asset returns was
investigated much earlier by Dr{\~a}gulescu and Yakovenko
\cite{Dragulescu2002}).  In \cite{Lanoiselee2018}, we proposed
to model the stochastic diffusivity of a particle by a Feller process
\cite{Feller1951}, also known as the square root process or the
Cox-Ingersoll-Ross process \cite{Cox1985}:
\begin{equation}  \label{eq:Feller}
dD_t = \frac{1}{\tau}(\bar{D}-D_t)dt + \sigma \sqrt{2D_t}\, dW_t . 
\end{equation}
The diffusivity $D_t$ randomly walks around its mean value $\bar{D}$
due to rapid fluctuations of the medium modeled by the standard white
noise $dW_t$.  The two other parameters of the model characterize the
strength of these fluctuations ($\sigma$) and the time scale of medium
re-arrangements ($\tau$).  For a particle moving in the
$d$-dimensional space $\R^d$ free of reactive targets and inert
obstacles, we derived the full propagator $P(\x,D,t|\x_0,D_0)$, i.e.,
the probability density for a particle started at $\x_0$ with the
initial diffusivity $D_0$ to be at $\x$ with the diffusivity $D$ at a
later time $t$.  When the control dimensionless parameter $\nu =
\bar{D}/(\tau\sigma^2)$ is integer, the Feller process
(\ref{eq:Feller}) is equivalent to the square of an $\nu$-dimensional
Ornstein-Uhlenbeck process used for modeling the stochastic
diffusivity in \cite{Jain2016,Jain2017,Chechkin2017}, and our model is
thus reduced to the former one.  However, integer values of $\nu$
correspond to a weak disorder.  In fact, the parameter $1/\nu$
characterizes the disorder strength, i.e., how broad is the
distribution of random variations of the diffusivity in a
heterogeneous medium.  This can be seen by rescaling the diffusivity
$D_t$ by $\bar{D}$ and the time $t$ by $\tau$ in
Eq. (\ref{eq:Feller}), in which case the factor $\sqrt{1/\nu}$ appears
in front of the fluctuation term (see also discussion in
Sec. \SIgeneral of the Method section).  As a consequence, our
extension to any real positive $\nu$ and, in particular, to the range
$0 < \nu < 1$ that was inaccessible in former works, brought
conceptually new features to the annealed model of heterogeneous
diffusion.


The above works were devoted to the dynamics itself (MSD scaling, weak
ergodicity breaking, non-Gaussian behavior of the propagator, etc.),
with no chemical reaction involved.  The notable exception is the work
by Jain and Sebastian \cite{Jain2016b}, in which the survival
probability in crowded re-arranging spherical domains was derived.
While some first-passage time problems and related reaction kinetics
in static disordered media have been addressed
\cite{Bouchaud1990,Havlin2002,Weiss1986,Hernandez1990a,Hernandez1990b,Budini2018},
most former studies were focused on the mean FPT and reaction rates.

In this letter, we couple heterogeneous diffusion to chemical
reactions in a medium containing perfectly reactive targets and inert
obstacles.  We describe a general mathematical framework to translate
many results for ordinary homogeneous Brownian motion to heterogeneous
diffusion.  In particular, we derive general spectral decompositions
of the full and marginal propagators, the survival probability, the
probability density function of the first-passage time to a reaction
event, and the macroscopic reaction rate of diffusion-limited
reactions.  We show how the dynamic disorder broadens the probability
density and increases the likelihood of both short and long
trajectories to reactive targets.  In other words, while the reaction
process is slowed down on average, some molecules can reach the
reactive targets much faster than via homogeneous diffusion.  We
discuss biological implications of this counter-intuitive finding,
further perspectives and open problems.

\section*{RESULTS}

\subsection*{Heterogeneous diffusion toward reactive targets}

Let us consider a particle diffusing in a fixed volume $\Omega \subset
\R^d$ outside an arbitrary configuration of immobile perfectly
reactive targets and inert obstacles.  The stochastic diffusivity
$D_t$, modeled by the Feller process (\ref{eq:Feller}), represents the
dynamic disorder due to rapid re-arrangements of the medium.  The
particle reacts upon the first encounter with any target and thus
disappears, being chemically transformed into another species.  This
is a standard scheme for most catalytic reactions.  In turn, inert
obstacles or impenetrable walls just hinder the motion of the particle
or confine it in a prescribed spatial region (e.g., inside a living
cell).  For any bounded domain $\Omega$ (e.g., the cytoplasm confined
by the plasma membrane), we obtain the spectral decomposition for the
full propagator $P(\x,D,t|\x_0,D_0)$ by solving the Fokker-Planck
equation (see Sec. \SIgeneral of the Method section).  As the
instantaneous diffusivities $D_0$ and $D$ are hard to access from
experiments, we focus throughout this letter on the more common
marginal propagator $P(\x,t|\x_0)$, which is obtained by averaging
$P(\x,D,t|\x_0,D_0)$ over the diffusivity $D$ at time $t$ and over the
initial diffusivity $D_0$ taken from its stationary distribution.  We
show in Sec. \SIgeneral of the Method section that the propagator
$P(\x,t|\x_0)$ admits a general spectral decomposition
\begin{equation}  \label{eq:main}
P(\x,t|\x_0) = \sum\limits_{n=1}^\infty u_n(\x) \, u_n(\x_0) \, \Upsilon(t; \lambda_n) ,
\end{equation}
where $\lambda_n$ and $u_n$ are the eigenvalues and the
$L_2$-normalized eigenfunctions of the Laplace operator $\Delta$ in
$\Omega \subset \R^d$, verifying $\Delta u_n + \lambda_n u_n = 0$,
subject to absorbing (Dirichlet) and reflecting (Neumann) boundary
conditions on the surfaces of targets and obstacles, respectively, and
\begin{equation}  \label{eq:Ups}
\Upsilon(t; \lambda) = \left( \frac{4\omega e^{-(\omega-1) t/(2\tau)}}{(\omega+1)^2 - (\omega-1)^2 e^{-\omega t/\tau} } \right)^\nu ,
\end{equation}
with $\omega = \sqrt{1 + 4\sigma^2\tau^2 \lambda}$.  Our setting and
derivation are much more general than that by Jain and Sebastian who
obtained a similar spectral decomposition for a disk with a perfectly
reactive boundary for diffusing diffusivity modeled by an
$\nu$-dimensional Ornstein-Uhlenbeck process \cite{Jain2016b}.  When
either dynamic re-arrangements of the medium are too fast ($\tau
\to 0$) or its fluctuations are too small ($\sigma \to 0$), the
diffusivity is constant, $D_t = \bar{D}$, Eq. (\ref{eq:Ups}) is
reduced to $\Upsilon_{\rm hom} = \exp(-\bar{D} t\lambda)$, and one
recovers the standard spectral decomposition of the propagator for
homogeneous diffusion \cite{Gardiner}.  While the dynamic disorder is
incorporated in Eq. (\ref{eq:main}) via the explicitly known function
$\Upsilon(t; \lambda)$, the structure of the confining domain and its
reactive properties are fully ``encoded'' by the Laplacian eigenmodes,
$\lambda_n$ and $u_n$ \cite{Grebenkov2013}.  The function
$\Upsilon(t;\lambda_n)$ couples, via the expression for $\omega$, the
geometric length scales $\lambda_n^{-1/2}$ of the reactive medium to
$\sigma\tau$, which can thus be understood as the disorder length
scale.

\subsection*{First-passage times to a reaction event}

The propagator is the essential ingredient for describing
diffusion-limited reactions.  In particular, the integral of the
propagator $P(\x,t|\x_0)$ over the arrival point $\x$ yields the
survival probability of a particle inside the domain, from which the
time derivative gives the probability density function of the
first-passage time to perfectly reactive targets on the boundary
$\pa$:
\begin{equation}  \label{eq:rho_main}
\rho(t|\x_0) = - \sum\limits_{n=1}^\infty  u_n(\x_0) \, \Upsilon'(t; \lambda_n) \int\limits_\Omega d\x \, u_n(\x)  ,
\end{equation}
where prime denotes the time derivative, $\Upsilon'(t; \lambda) =
\frac{\partial}{\partial t} \Upsilon(t; \lambda)$, which is known
explicitly from Eq. (\ref{eq:Ups}):
\begin{equation}  \label{eq:dP}  
\Upsilon'(t;\lambda) = - \frac{\nu}{2\tau}
\left(\omega-1 + \frac{2\omega \bigl(\frac{\omega-1}{\omega+1}\bigr)^2 e^{-\omega t/\tau}}{1 - \bigl(\frac{\omega-1}{\omega+1}\bigr)^2
e^{-\omega t/\tau} }\right) \Upsilon(t; \lambda).
\end{equation}
The probability density $\rho(t|\x_0)$ is the likelihood for the
reaction event to occur at a given time $t$.  Setting appropriate
Dirichlet-Neumann boundary conditions, one can describe, for instance,
the distribution of the reaction time on catalytic germs in a chemical
reactor, or the distribution of the first exit time from a confining
domain through ``holes'' on the boundary (e.g., through water or ion
channels on the plasma membrane of a living cell).  More generally,
this formalism allows one to ``translate'' many first-passage results
known for homogeneous diffusion to heterogeneous one and thus to
investigate the impact of the dynamic disorder onto heterogeneous
catalysis, diffusive search for multiple targets and escape problems
\cite{Redner,Metzler,Benichou2014,Holcman2013,Holcman2014,Grebenkov2017b}.

When the number of particles is large, multiple reaction events occur
at different times, and the overall chemical production can be
accurately characterized by the mean FPT or by the macroscopic
reaction rate $J(t)$, i.e., the overall flux of diffusing particles
onto the reactive target at time $t$.  As $\rho(t|\x_0)$ can be
interpreted as the probability flux onto the target at time $t$ for a
single particle started at $\x_0$ at time $0$, the overall flux $J(t)$
is obtained by superimposing these contributions.  If there are many
independent diffusing particles with a prescribed initial
concentration $c_0(\x_0)$, each contribution $\rho(t|\x_0)$ is
weighted by the number of particles at $\x_0$ (i.e., by
$c_0(\x_0)d\x_0$) that yields
\begin{equation}  \label{eq:J_main}
J(t) = - \sum\limits_{n=1}^\infty  \Upsilon'(t; \lambda_n) 
\int\limits_\Omega d\x_0 \, c_0(\x_0) \, u_n(\x_0)   \int\limits_\Omega d\x \, u_n(\x) .
\end{equation}
However, many cellular processes are triggered by the arrival of one
or few molecules onto the target (e.g., a receptor), and the number of
such molecules inside the cell is small.  In this case, the mean FPT
and the macroscopic rate $J(t)$ are not representative, and the full
distribution of the first-passage time is needed 
\cite{Grebenkov2018}.  Equation (\ref{eq:rho_main}) is thus the
crucial step to understand the reaction kinetics in re-arranging
heterogeneous media.  In the following, we focus on the probability
density $\rho(t|\x_0)$, bearing in mind straightforward extensions to
the reaction rate (its behavior is illustrated in the Method section).

\subsection*{Fast and slow arrivals to reactive targets}

The probability density $\rho(t|\x_0)$ can span over many orders of
magnitude in time so that two reaction times in the same medium can be
dramatically different.  In order to grasp such a broadness of
reaction times, it is instructive to look at reaction events that
occur at short and long times after the particle release.

The long-time behavior of the probability density function is
determined by the smallest eigenvalue $\lambda_1 > 0$:
\begin{equation}   \label{eq:rho_long} 
\rho(t|\x_0) \propto u_1(\x_0) \biggl(\frac{4\omega_1}{(\omega_1+1)^2}\biggr)^\nu \, \exp\biggl(- \frac{2}{1+\omega_1}\, \bar{D} t \lambda_1\biggr) ,
\end{equation}
with $\omega_1 = \sqrt{1 + 4\sigma^2\tau^2 \lambda_1}$.  This right
tail of the probability density characterizes long trajectories to
reactive targets.  A diffusing particle fully explores the confining
domain and thus almost looses track of the starting point $\x_0$ that
affects only a prefactor via the eigenfunction $u_1(\x_0)$.  The
asymptotic behavior is therefore mainly determined by the eigenvalue
$\lambda_1$ which in general exhibits an intricate dependence on the
geometry of the confining domain and on the configuration of reactive
targets \cite{Grebenkov2013,Holcman2014}.  The exponential decay of
the probability density function resembles that for homogeneous
diffusion with the mean diffusivity $\bar{D}$, but the decay rate is
decreased by the factor $(1+\omega_1)/2 \geq 1$.  When the disorder
length scale $\sigma\tau$ is much smaller than the largest geometric
scale $\lambda_1^{-1/2}$, then $\omega_1 \approx 1$, and one recovers
the long-time behavior known for homogeneous diffusion.  In this
limit, the particle has enough time to probe various diffusivities and
to average out the disorder.  In the opposite limit of a long-range
disorder, $\sigma \tau \gg \lambda_1^{-1/2}$, the decay rate in the
exponential function is greatly reduced by the factor $\sigma\tau
\sqrt{\lambda_1} \gg 1$, and thus the right tail of the
probability density is increased.  In particular, the mean FPT to a
reactive target, which is essentially determined by the exponential
tail, is increased by the factor $\sigma\tau \sqrt{\lambda_1}$.  We
conclude that the dynamic disorder slows down, on average, the
reaction kinetics and search by a single particle.

The short-time behavior of the probability density function of the
first-passage time to a perfectly reactive region $\Gamma$ of the
boundary is deduced in Sec. \SISS of the Method section:
\begin{equation}  \label{eq:rho_short}   
\rho(t|\x_0) \propto  t^{-1}  \, \biggl(\delta \sqrt{\nu/(\bar{D} t)}\biggr)^{\nu} \, \exp\biggl(- \delta \sqrt{\nu/(\bar{D}t)}\biggr)  ,
\end{equation}
where $\delta$ is the distance between the starting point $\x_0$ and
the reactive region $\Gamma$.  This relation, which is valid for $t
\ll \min\{\tau, \nu \delta^2/\bar{D}\}$, characterizes short, almost
direct trajectories to reactive targets, along which the diffusivity
remained almost constant.  Looking at the argument of the exponential
function in Eq. (\ref{eq:rho_short}), one can appreciate a dramatic
effect of heterogeneous diffusion at short times; in particular, the
decay of the probability density function for homogeneous diffusion is
much faster:
\begin{equation}  \label{eq:rhoBM_short}
\rho_{\rm hom}(t|\x_0) \propto t^{-1} \bigl(\delta/\sqrt{\bar{D} t}\bigr) \, \exp\bigl(-\delta^2/(4\bar{D} t)\bigr).  
\end{equation}
As a consequence, rapid arrivals of a particle to the reactive region
are much more probable for heterogeneous diffusion.  In other words,
the dynamic character of the disorder allows for larger diffusivities
and is thus beneficial for a faster arrival to the target by a single
particle, in spite the longer mean FPT.  The most probable
first-passage time, at which the probability density function reaches
its maximum, $(\partial \rho(t|\x_0)/\partial t)|_{t_{\rm mp}} = 0$,
can be estimated from Eq. (\ref{eq:rho_short}) as $t_{\rm mp}
\approx (1 + 5/(2\nu))^{-1} \, \delta^2/\bar{D}$.  As
expected, the most probable FPT is proportional to $\delta^2/\bar{D}$
as for Brownian motion but the prefactor is controlled by the disorder
strength $1/\nu$.  In particular, the most probable FPT goes to $0$ as
the disorder strength $1/\nu$ increases.  The distance to the target,
$\delta$, is the only relevant geometric length in the short-time
regime, which is thus very sensitive to the starting point $\x_0$.

\subsection*{Respective roles of the disorder strength and scale}

While the above asymptotic relations are universal, the
functional form of the probability density $\rho(t|\x_0)$ depends on
the shape of the confining domain and its reactive properties.  In
spite of intensive studies over the past decades
\cite{Benichou2010,Havlin2002,Sokolov2012,Hofling2013,Redner,Metzler,Grebenkov2005,Levitz2006,Grebenkov2016},
the strong impact of the geometric complexity onto first-passage times
and chemical reactions is not fully understood even for homogeneous
diffusion.  In order to decouple the geometric aspects from the
dynamic disorder, we consider as an illustrative example heterogeneous
diffusion in a simple yet emblematic domain -- a ball.  This is a very
common model of confinement that was in the scope of many former
theoretical studies.  Since the radius $R$ of the ball is the only
geometric scale of the domain, one can focus exclusively on the impact
of the dynamic disorder.  The substitution of the explicit form of
Laplacian eigenmodes \cite{Carslaw,Grebenkov2013} into
Eq. (\ref{eq:rho_main}) yields
\begin{equation}  \label{eq:rho_sphere}
\rho(t|\x_0) = 2 \sum\limits_{n=1}^\infty (-1)^n \underbrace{\frac{\sin(\pi n \|\x_0\|/R)}{\pi n \|\x_0\|/R}}_{ = 1 ~ \textrm{if} ~ \x_0 = 0} 
\,  \Upsilon'(t; \pi^2 n^2/R^2)  ,
\end{equation}
where $\|\x_0\|$ is the radial coordinate of the starting point
$\x_0$.  Moreover, we consider the first-passage time to the boundary
of a ball from its center, $\x_0 = \bm{0}$, to fix the distance to the
target: $\delta = R$.  Given that the parameters $R$ and $R^2/\bar{D}$
fix the length and time scales for homogeneous diffusion, we
investigate the impact of the two other parameters of the dynamic
disorder.  The asymptotic relations (\ref{eq:rho_long},
\ref{eq:rho_short}) suggest that the proper dimensionless parameters
of the model are the disorder scale $\mu = \sigma\tau/R$ (which
compares the spatial extent of the disorder to the size of the
domain), and the disorder strength $1/\nu = \tau
\sigma^2/\bar{D}$.

Figure \ref{fig:rhot_int} compares the exact solution
(\ref{eq:rho_sphere}) for heterogeneous diffusion, with
$\Upsilon'(t;\lambda)$ from Eq. (\ref{eq:dP}), and for homogeneous
diffusion with mean diffusivity $\bar{D}$ and $\Upsilon'_{\rm
hom}(t;\lambda) = - \bar{D} \lambda \, \exp(-\bar{D}t\lambda)$.  We
explore the parameters space $(\mu,1/\nu)$ in two complementary ways.
In the top panels (a,b,c), we fix three values of the disorder scale
$\mu$ ($10^{-1}$, $1$, and $10$) and range ``continuously'' the
disorder strength $1/\nu$ from $10^{-1}$ to $10^1$.  When the disorder
scale is small ($\mu = 0.1$), the particle travels enough distance to
the reactive boundary to average out stochastic diffusivities.  As a
consequence, the long-time behavior of the probability density (its
right tail) is close to that of homogeneous diffusion with the mean
$\bar{D}$, regardless the disorder strength $1/\nu$ in the considered
range.  At larger disorder scales ($\mu = 1$ and $\mu = 10$),
deviations from homogeneous diffusion at long times become
progressively stronger.  An increase of the disorder strength $1/\nu$
leads to progressive broadening of the distribution.  In particular,
the short-time tail of the probability density function is shifted to
the left, increasing thus chances of reaching the target at short
times.  In contrast, the short-time behavior remains almost unaffected
by the scale $\mu$, when $\mu$ is not too small (compare cases $\mu =
1$ and $\mu = 10$).  This is more clearly seen in the bottom panels
(d,e,f), which show $\rho(t|0)$ for three fixed values $1/\nu$
($10^{-1}$, $1$, and $10$) and numerous values of $\mu$ ranging from
$10^{-1}$ to $10^1$.  The left short-time tail is almost independent
of $\mu$ and controlled by $1/\nu$, in agreement with the short-time
asymptotic relation (\ref{eq:rho_short}).  In turn, the right tail is
affected by both $\mu$ and $\nu$, see Eq. (\ref{eq:rho_long}).  As the
disorder weakens ($1/\nu \to 0$ with fixed $\mu$), the probability
density $\rho(t|0)$ approaches $\rho_{\rm hom}(t|0)$ for homogeneous
diffusion.  In turn, the short-time tail of $\rho(t|0)$ exhibits
deviations from $\rho_{\rm hom}(t|0)$ in the other limit $\mu\to 0$
(with fixed $\nu$), as discussed in Sec. \SIlimit of the Method
section.

\section*{DISCUSSION}

The discovered broadening of the distribution and increase of its both
short- and long-time tails by dynamic disorder are generic and valid
for bounded domains beyond balls.  Moreover, our study can be extended
to unbounded domains, for which the analysis becomes more subtle
because the spectrum of the Laplace operator is not discrete anymore.
In Sec. \SIunbounded~ of the Method section, we provide the explicit
representations of the propagator, the survival probability, the
probability density, and the macroscopic reaction rate for two
unbounded domains: a half-space with a perfectly reactive hyperplane
and the exterior of a perfectly reactive ball.  In both cases, we show
that the long-time behavior of heterogeneous diffusion approaches that
of the homogeneous one: the particle has enough time to average out
the dynamic disorder, whatever its length scale $\sigma\tau$ (this is
equivalent to $\mu = 0$).  In particular, we retrieve the Smoluchowski
diffusion-limited reaction rate for a spherical target as time goes to
infinity \cite{Smoluchowski1917}, while the approach to this
stationary limit is moderately affected by the dynamic disorder. 

\begin{figure}[t!]
\begin{center}  
\includegraphics[width=85mm]{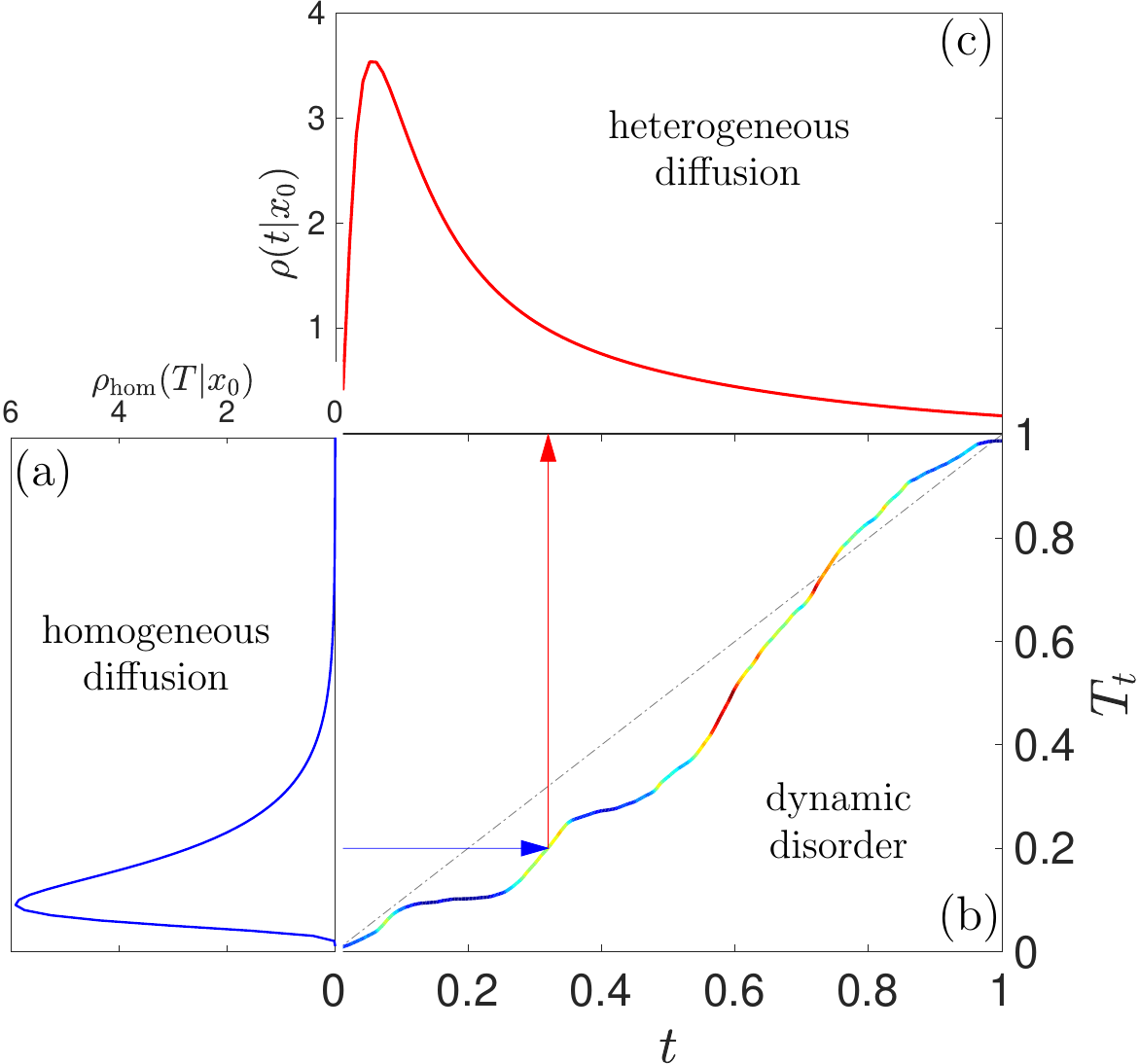}
\end{center}
\caption{
{\bf Illustration of the subordination concept.}  The first-passage
time to the reactive target is understood as the moment of the first
crossing of a random barrier by the integrated diffusivity $T_t$.
{\bf (a)} The geometric structure and reactive properties of the
medium determine the probability density function $\rho_{\rm
hom}(T|\x_0)$ of the first-passage time $\T_{\rm hom}$ to the reactive
target by homogeneous diffusion.  This FPT is the ``duration'' of a
random Brownian path to the reactive target that sets the barrier to
the integrated diffusivity $T_t$ (horizontal blue arrow).  The
randomness of such Brownian paths results from thermal fluctuations.
{\bf (b)} Rapid re-arrangements of the medium lead to a random
realization of the ``internal time'' $T_t$, obtained by integrating
the stochastic diffusivity $D_t$ shown in Fig. \ref{fig:actin}(c);
colors are reproduced from that figure, ranging from dark blue (low
diffusivity) to dark red (high diffusivity).  The random moment $t$
(shown by vertical red arrow) when $T_t$ crosses the random barrier
$\T_{\rm hom}$ is the first-passage time to the reactive target by
heterogeneous diffusion.  {\bf (c)} The probability density
$\rho(t|\x_0)$ of this FPT is obtained by averaging the density
$q(t;\T_{\rm hom})$ over the distribution of $\T_{\rm hom}$ given by
$\rho_{\rm hom}(T|\x_0)$.  The broadening of $\rho(t|\x_0)$ is caused
by superimposing two sources of randomness in heterogeneous diffusion:
thermal fluctuations (as in $\rho_{\rm hom}(T|\x_0)$) and medium
re-arrangements.  Arbitrary units are used for this illustrative
picture.}
\label{fig:subord}
\end{figure}

So far, we investigated the impact of the dynamic disorder onto
chemical reactions for the particular model (\ref{eq:Feller}) of
diffusing diffusivity.  But, the derived spectral decompositions
(\ref{eq:main}, \ref{eq:rho_main}) turn out to be much more general
and can couple the geometric structure of the reactive confining
domain $\Omega$ to an arbitrary model of the dynamic disorder
represented via the function $\Upsilon(t;\lambda)$.  In fact, there
are two independent sources of randomness in the annealed model of
heterogeneous diffusion: thermal fluctuations that result from local
interactions of the medium with a diffusing particle and drive its
stochastic motion, and rapid re-arrangements of the medium that change
the ``amplitude'' of the local interactions via the stochastic
diffusivity.  The Laplacian eigenmodes determine the statistics of all
possible random paths of a particle in a homogeneous medium due to
thermal fluctuations.  In turn, the diffusing diffusivity $D_t$
affects only the ``speed'' at which the particle moves along a
randomly chosen path (Fig. \ref{fig:actin}).  This is the idea of
subordination when the integrated diffusivity, $T_t =
\int\nolimits_0^t dt' D_{t'}$, is considered as the ``internal time''
of a homogeneous process \cite{Chechkin2017}.  If the propagator
$P_{\rm hom}(\x,T|\x_0)$ of the homogeneous process with a fixed
internal time $T$ is known, then the propagator for the subordinated
heterogeneous process, in which $T = T_t$ is a random variable, is
obtained by averaging $P_{\rm hom}(\x,T_t|\x_0)$ with the probability
density function $Q(t;T)$ of the integrated diffusivity $T_t$:
\begin{equation}
P(\x,t|\x_0) = \int\limits_0^\infty dT \, Q(t;T) \, P_{\rm hom}(\x,T|\x_0).
\end{equation}
This relation naturally couples two sources of randomness: thermal
fluctuations (determining $P_{\rm hom}(\x,T|\x_0)$) and the dynamic
disorder (determining $Q(t;T)$).  For a homogeneous diffusion in a
bounded medium with reactive targets, $P_{\rm hom}(\x,T|\x_0)$ admits
the standard spectral decomposition \cite{Gardiner}, from which
Eq. (\ref{eq:main}) follows, with
\begin{equation}  \label{eq:QT_def}
\Upsilon(t;\lambda) = \int\limits_0^\infty dT \, e^{-\lambda T} \, Q(t;T)
\end{equation}
being the Laplace transform of the probability density function
$Q(t;T)$.  In the same vein, the subordination form for the
first-passage time density $\rho(t|\x_0)$ is derived in
Sec. \SIsubordination~ of the Method section:
\begin{equation}
\rho(t|\x_0) = \int\limits_0^\infty dT \, q(t;T) \, \rho_{\rm hom}(T|\x_0) ,
\end{equation}
where $q(t;T)$ is the probability density function of the first
crossing time of a prescribed barrier at $T$ by the integrated
diffusivity $T_t$ (the density $q(t;T)$ is also directly related to
$Q(t;T)$ and $\Upsilon(t;\lambda)$, see Sec. \SIsubordination~ of the
Method section).  This subordination concept is illustrated by
Fig. \ref{fig:subord}.

In this light, the Feller process (\ref{eq:Feller}) for diffusivity
$D_t$ can be replaced by another process to reproduce the desired
features of dynamic heterogeneous media.  In the simplest case when
the time-dependent diffusivity $D_t$ is deterministically prescribed,
$T_t$ is not random, so that $Q(t;T) = \delta(T - T_t)$ and thus
$\Upsilon(t;\lambda) = \exp\bigl(-\lambda T_t)$.  When the particle
undergoes a continuous-time random walk with long stalling periods
characterized by an anomalous waiting exponent $0 < \alpha < 1$
\cite{Metzler2000}, one gets $\Upsilon(t;\lambda) =
E_\alpha(- D_\alpha t^\alpha \lambda)$, where $E_\alpha(z)$ is the
Mittag-Leffler function, and $D_\alpha$ is the (constant) generalized
diffusion coefficient \cite{Fa2005b,Grebenkov2010a}.  One can also
consider L\'evy-noise-driven processes to model diffusivity with heavy
tails \cite{Jain2017a}, geometric Brownian motion to get a
nonstationary evolution, or a customized stochastic process to
produce the desired distribution of the stationary diffusivity
\cite{Sposini2018}.  Once the function $\Upsilon(t;\lambda)$ is
computed for the chosen diffusivity model, the coupling to the spatial
dynamics of the particle, the related first-passage phenomena, and the
consequent reaction kinetics are immediately accessible via the
spectral decomposition (\ref{eq:main}).  We stress however that the
subordination does not provide the full propagator
$P(\x,D,t|\x_0,D_0)$ but only the marginal propagator $P(\x,t|\x_0)$.

This letter was focused on diffusion-limited reactions because
the related first-passage statistics are essential for characterizing
the diffusive transport toward the targets.  However, many
(bio)chemical reactions involve other ``ingredients'' such as active
transport by motor proteins, bulk reactivity, partially reactive
targets, reversible association-dissociation processes and re-binding
effects, collective search by multiple particles and the associated
(anti-)cooperativity effects, surface diffusion and intermittence, to
name but a few.  These effects have been progressively incorporated
into the theory of homogeneous diffusion-controlled reactions during
the past century since the Smoluchowski's seminal paper
\cite{Smoluchowski1917}.  Some of these ingredients can be immediately
implemented into our formalism.  For instance, the Laplace operator
governing passive diffusion can be replaced by more general
Fokker-Planck operators accounting for an external potential or a
drift, allowing one to model active transport in dynamic heterogeneous
media such as the cytoplasm of living cells
\cite{Bressloff2013}.  In turn, the inclusion of some other ingredients
remains challenging and requires future investigations.  For instance,
the macroscopic description of homogeneous diffusion in a medium with
partially reactive targets employs the Robin boundary condition that
equates the diffusive flux density $-D_0
\frac{\partial}{\partial n} P(\x,t|\x_0)$ toward the target to the
reactive flux density $\kappa P(\x,t|\x_0)$ on the target, the
reactivity $\kappa$ characterizing the efficiency of reaction (and
$\frac{\partial}{\partial n}$ being the normal derivative).  An
extension of this condition to heterogeneous diffusion with random
diffusivity $D_t$ instead of $D_0$ does not seem possible for the
marginal propagator $P(\x,t|\x_0)$ and requires considering the full
propagator $P(\x,D,t|\x_0,D_0)$.  The apparent simplicity of the
implementation of the dynamic disorder into the realm of homogeneous
diffusions via the function $\Upsilon(t;\lambda)$ is thus deceptive,
and the implementation of partial reactivity and some other mechanisms
for heterogeneous diffusion raises open mathematical questions.

Another important perspective consists in developing new statistical
tools, based on the proposed formalism, to distinguish the impact of
the dynamic disorder from other intracellular features (such as
visco-elasticity and overcrowding), to identify proper models of
diffusing diffusivity from experimental single-particle trajectories,
and to infer the parameters of that models.  In particular, molecular
dynamics simulations could help identifying such models from
microscopic principles.  In turn, Monte Carlo and finite elements
methods allow one to further investigate the role of multiple
geometric length scales onto the reaction kinetics in complex
geometric confinements.

In summary, we discussed the impact of spatio-temporal disorder of
dynamic heterogeneous media onto diffusion-limited reactions, bearing
in mind applications to intracellular reactions.  A conventional way
of tackling such problems would consist in modeling the whole
dynamically re-arranging medium by means of molecular dynamics
simulations.  However, a living cell is a very complex system in which
a vast number of particles, from water, ions, proteins, actin
filaments and microtubules to large organelles such as vesicles and
mitochondria, interact to each other, all being confined between the
nucleus and the plasma membrane.  Even though molecular dynamics
simulations of the intracellular dynamics become more and more
accurate and large-scale \cite{McGuffee2010,Yu2016}, understanding the
respective impacts of different cellular mechanisms and processes
remains challenging.  Theoretical approaches offer a complementary
insight by focusing on a particular feature of the intracellular
dynamics and ignoring its other aspects.  For instance, generalized
Langevin equations with memory kernels were invoked to capture
visco-elastic properties of the cytoplasm and the related long-time
corrections, whereas continuous-time random walk can model molecular
caging in an overcrowded environment.  Combining such individual
mechanisms as elementary pieces, one aims at reconstructing, step by
step, the whole mosaic of a cell life.  Here, we added a new puzzle
element by investigating the effects related to dynamic
re-arrangements of the intracellular medium due to, e.g., actin waves
or microtubule movement \cite{Kulic2008,Allard2013}.  We greatly
simplified the problem by modeling the impact of the medium onto the
particle via diffusing diffusivity and thus reducing irrelevant
degrees of freedom.  The developed theoretical framework revealed that
dynamic heterogeneities can actually be beneficial for many
biochemical processes in living cells which are triggered by a single
molecule \cite{Xie2006,Xie2011}.  More generally, we provided a
mathematical ground to advance understanding and modeling of
intracellular dynamics to a new level, with potential biomedical and
pharmaceutical applications.

\newpage
\section*{METHOD}

\subsection*{1. Derivation of the propagator}

The derivation of the propagator in $\R^d$ from \cite{Lanoiselee2018}
can be generalized to an arbitrary bounded domain $\Omega \subset
\R^d$, in which the eigenvalue problem for the Laplace operator is
well defined \cite{Gardiner,Grebenkov2013}.  The probability density
$P(\x,D,t|\x_0,D_0)$ for a particle started from a point $\x_0$ with
the initial diffusivity $D_0$ to be at a point $\x$ with the
diffusivity $D$ at a later time $t$ satisfies the forward
Fokker-Planck equation in the It\^o convention:
\begin{equation}  \label{eq:FP}
\frac{\partial P}{\partial t} = \frac{1}{\tau} \frac{\partial}{\partial D} \bigl((D-\bar{D}) P\bigr) 
+ D\Delta P + \sigma^2 \frac{\partial^2}{\partial^2 D} (DP) ,
\end{equation}
subject to the initial condition $P(\x,D,t=0|\x_0,D_0) = \delta(\x -
\x_0) \delta(D - D_0)$ and an appropriate boundary condition on the
boundary $\pa$ of the domain $\Omega$.  While the Langevin equation
(\ref{eq:Feller}) automatically ensures the positivity of the
diffusivity $D_t$ \cite{Feller1951,Cox1985}, the Fokker-Planck
equation needs an additional condition at the boundary $D = 0$ in the
phase space $(\x,D)$.  As discussed in detail in
\cite{Lanoiselee2018}, two standard conditions are often employed: the
absorbing condition $P(\x,D = 0,t|\x_0,D_0) = 0$ and no flux condition
$J_D = - \bigl(\frac{1}{\tau}(D - \bar{D})+
\sigma^2 \frac{\partial}{\partial D} DP\bigr)|_{D=0} = 0$.  The former
condition implies that random trajectories in the phase space $(\x,D)$
stop after hitting the boundary $D = 0$: once the diffusivity $D_t$
reaches $0$, it gets stuck in this state.  As this situation is
unphysical, we choose the second condition that ensures the strict
positivity of the diffusivity \cite{Lanoiselee2018,Gan2015}.  We also
impose the regularity condition $P(\x,D,t|\x_0,D_0) \to 0$ as
$D\to\infty$.

We sketch the main steps of the derivation.  First, one applies the
Laplace transform with respect to $D \geq 0$:
\begin{equation}
\tilde{P}(\x,s,t|\x_0,D_0) = \int\limits_0^\infty dD \, e^{-sD} \, P(\x,D,t|\x_0,D_0),
\end{equation}
to transform Eq. (\ref{eq:FP}) to
\begin{equation}  \label{eq:FP2}
\frac{\partial \tilde P}{\partial t} + \bigl(\sigma^2 s^2 + s/\tau + \Delta) 
\frac{\partial}{\partial s} \tilde P = - \frac{\bar{D} s}{\tau} \tilde P ,
\end{equation}
where we used no flux condition at $D = 0$.  We decompose $\tilde P$
on the complete basis of orthonormal Laplacian eigenfunctions,
verifying $\Delta u_n(\x) + \lambda_n u_n(\x) = 0$ in $\Omega$, with
the desired boundary condition on $\pa$, and $\lambda_n$ being the
eigenvalues enumerated by $n = 1,2,\ldots$ in an increasing order.
Moreover, the orthogonality of eigenfunctions allows one to search the
propagator $\tilde P$ in the form
\begin{equation}  \label{eq:tildeP_eigen}
\tilde P(\x,s,t|\x_0,D_0) = \sum\limits_{n=1}^\infty u_n(\x) \, u_n(\x_0) \, \tilde p(\lambda_n,s,t|D_0) .
\end{equation}
Substitution of this form into Eq. (\ref{eq:FP2}) yields a first-order
differential equation for the unknown function $\tilde
p(\lambda,s,t|D_0)$:
\begin{equation}
\frac{\partial \tilde p}{\partial t} + \bigl(\sigma^2 s^2 + s/\tau - \lambda\bigr) \frac{\partial \tilde p}{\partial s}
= - \frac{\bar{D} s}{\tau} \tilde p ,
\end{equation}
subject to the initial condition $\tilde p(\lambda,s,t=0|D_0) =
e^{-sD_0}$.  The above equation was solved in
\cite{Lanoiselee2018} by the method of characteristics:
\begin{eqnarray}   \label{eq:complete_solution}
&& \tilde p(\lambda,s,t|D_0) = F(D_0,s) \, e^{ -\nu (\omega-1) t/(2\tau)}\\  \nonumber
&&\times\left(\frac{\sigma^2\tau}{\omega}\left[\left(s+\frac{1+\omega}{2\sigma^2\tau}\right)-
\left(s+\frac{1-\omega}{2\sigma^2\tau}\right) e^{-\omega t/\tau}\right]\right)^{-\nu},
\end{eqnarray}
where
\begin{align}
F(D_0,s) &= \exp\left[\frac{D_0}{2\sigma^2\tau}\left(1+\omega-\frac{2\omega}{1-\xi
e^{-\omega t/\tau}}\right)\right], \\
\xi &= 1-\frac{2\omega}{1+\omega+2\sigma^2\tau s} \,,  \\
\label{eq:omega}
\omega &= \sqrt{1+4\sigma^2\tau^2 \lambda} \,.  
\end{align}
These relations provide the exact formula (\ref{eq:tildeP_eigen}) for
the propagator in the Laplace domain with respect to the diffusivity
$D$.  The inverse Laplace transform is in general needed to get the
propagator $P(\x,D,t|\x_0,D_0)$.  

As the diffusivity $D$ at time $t$ is not relevant for most
applications, one can focus on the marginal distribution of the
position $\x$ by integrating over $D$, which is obtained by setting $s
= 0$ in $\tilde p(q,s,t|D_0)$:
\begin{equation}  \label{eq:Pmarg}
P(\x,t|\x_0,D_0) = \sum\limits_{n=1}^\infty u_n(\x) \, u_n(\x_0) \, \Upsilon(t; \lambda_n|D_0) ,
\end{equation}
where 
\begin{eqnarray}    \label{eq:Upsilon_D0}
&& \Upsilon(t; \lambda|D_0) = \biggl(\frac{2\omega e^{-(\omega-1)t/(2\tau)}}{\omega+1 + (\omega-1)e^{-\omega t/\tau}}\biggr)^\nu \\
\nonumber
&& \times \exp\biggl(\frac{D_0 (\omega+1)}{2\sigma^2\tau} \biggl(1 - \frac{2\omega}{\omega+1 + (\omega-1)e^{-\omega t/\tau}}\biggr)\biggr).
\end{eqnarray}
The marginal distribution (\ref{eq:Pmarg}) is fully explicit in terms
of the time dependence.

When the medium is rapidly fluctuating, it is difficult to control the
initial diffusivity $D_0$.  Since the Feller process (\ref{eq:Feller})
for the stochastic diffusivity $D_t$ is stationary, a random
``pickup'' of the initial diffusivity $D_0$ can be naturally realized
by using the stationary distribution of $D_t$ which is known to be the
Gamma distribution \cite{Feller1951,Lanoiselee2018}
\begin{equation}  \label{eq:Gamma}
\Pi(D) = \frac{\nu^\nu \, D^{\nu-1}}{\Gamma(\nu) \,\bar{D}^\nu} \, e^{-\nu D/\bar{D}} ,
\end{equation}
characterized by the scale $\bar{D}/\nu$ (with the mean $\bar{D}$) and
the shape parameter $\nu = \bar{D}/(\tau\sigma^2)$ (with $\Gamma(\nu)$
being the Euler gamma function).  We note that, from a physical point
of view, local diffusivities should be bounded by the diffusivity of
the particle in water, $D_{\rm max}$, which should thus provide a
finite cut-off of the distribution.  However, the mean diffusivity
$\bar{D}$ in the cytoplasm is much smaller than $D_{\rm max}$ so that
the probability of getting diffusivities larger than $D_{\rm max}$ is
exponentially small.  In other words, the exponential decay in
Eq. (\ref{eq:Gamma}) effectively substitutes the finite cut-off.

The $k$-th moment of the stationary diffusivity reads
\begin{equation}
\langle D^k \rangle = \begin{cases} \displaystyle \frac{\Gamma(\nu+k)}{\Gamma(\nu)\, \nu^k} \, \bar{D}^k \quad (k > -\nu),  \cr
\infty \hskip 19.5mm  (k \leq -\nu),  \end{cases}
\end{equation}
which is valid even for non-integer and negative $k$.  From this
relation, one can express the inverse of the shape parameter as
\begin{equation}  \label{eq:one_nu}
\frac{1}{\nu} = \frac{\mathrm{var}\{D\}}{\mathrm{mean}\{D\}^2} \,, 
\end{equation}
i.e., $1/\nu$ characterizes the strength of diffusivity heterogeneity:
larger $1/\nu$ corresponds to a broader distribution of stationary
diffusivities and thus to stronger disorder.  More generally, the role
of the parameter $1/\nu$ can be seen by rescaling the diffusivity
$D_t$ by its mean $\bar{D}$ and the time $t$ by $\tau$ in
Eq. (\ref{eq:Feller}) that gives
\begin{equation}  \label{eq:Feller_renorm}
d(D_t/\bar{D}) = (1 - D_t/\bar{D}) d(t/\tau) + \sqrt{1/\nu}\, \sqrt{2D_t/\bar{D}} \, dW_{t/\tau} .
\end{equation}
The factor $\sqrt{1/\nu}$ controls the amplitude of the fluctuation
term and thus the strength of the dynamic disorder.

The average of the propagator $P(\x,t|\x_0,D_0)$ over random
realizations of the initial diffusivity $D_0$, drawn from the Gamma
distribution (\ref{eq:Gamma}), yields the marginal propagator in a
bounded domain:
\begin{equation}
P(\x,t|\x_0) = \int\limits_0^\infty dD_0 \, \Pi(D_0) \, P(\x,t|\x_0,D_0) .
\end{equation}
Substitution of Eq. (\ref{eq:Pmarg}) into this relation implies the
spectral decomposition (\ref{eq:main}), with
\begin{equation}
\Upsilon(t; \lambda) = \int\limits_0^\infty dD_0 \, \Pi(D_0) \, \Upsilon(t; \lambda|D_0) .
\end{equation}
The computation of this integral yields Eq. (\ref{eq:Ups}).

The reactive properties of the boundary of the confining domain
$\Omega$ and its interaction with diffusing particles are introduced
via boundary conditions in a standard way \cite{Gardiner,Redner} and
fully captured by the Laplacian eigenmodes.  When the boundary is a
passive, impenetrable wall that constrains the particle inside a
bounded domain, Neumann boundary condition is imposed to ensure no
probability flux across the boundary: $\partial P/\partial n = 0$,
where $\partial/\partial n$ is the normal derivative.  In this case,
the particle is always present in the domain, and the normalization of
the propagator is preserved in time:
\begin{equation}
\int\limits_\Omega d\x \, P(\x,t|\x_0) = 1 .
\end{equation}
However, when the boundary contains holes, traps or reactive regions
that may kill, adsorb, transfer or transform the particle or modify
its state upon the first encounter, Dirichlet boundary condition is
imposed on these perfectly reactive parts of the boundary.  In this
case, the propagator $P(\x,t|\x_0)$ should be interpreted as the
probability density for a particle started at $\x_0$ to be found at
$\x$ at time $t$, without being destroyed or modified on its way.  As
a consequence, the normalization of the propagator is not preserved
and gradually decreases with time.  The above integral yields thus the
survival probability up to time $t$, $S(t|\x_0)$, for which
Eq. (\ref{eq:main}) implies
\begin{equation}
S(t|\x_0) = \sum\limits_{n=1}^\infty u_n(\x_0) \, \Upsilon(t; \lambda_n) \int\limits_\Omega d\x \, u_n(\x) .
\end{equation}
This quantity can also be understood as one minus the cumulative
distribution function (cdf) of the random first-passage time $\T$, at
which the particle reaches the target to be destroyed, chemically
transformed or modified on the reactive region: $S(t|\x_0) = 1 -
\P_{\x_0}\{ \T < t\}$.  In other words, $\T$ is the first-passage time
to a reaction event, whatever its microscopic mechanism is.  The time
derivative of the cdf gives the probability density function of this
first-passage time:
\begin{equation}
\rho(t|\x_0) = - \sum\limits_{n=1}^\infty u_n(\x_0) \, \Upsilon'(t; \lambda_n) \int\limits_\Omega d\x \, u_n(\x) ,
\end{equation}
where $\Upsilon'(t;\lambda)$, given explicitly by Eq. (\ref{eq:dP}),
denotes the time derivative of $\Upsilon(t;\lambda)$ from
Eq. (\ref{eq:Ups}).  Finally, the mean FPT can be obtained by
integrating $t \rho(t|\x_0)$ over $t$ from $0$ to $\infty$ that yields
\begin{equation}
\langle \T\rangle_{\x_0} = \sum\limits_{n=1}^\infty u_n(\x_0) \, \int\limits_\Omega d\x \, u_n(\x) \, \int\limits_0^\infty dt \, \Upsilon(t; \lambda_n).
\end{equation}
Note that the last integral can be expressed in terms of the Gauss
hypergeometric function as
\begin{eqnarray}
&& \int\limits_0^\infty dt \, \Upsilon(t; \lambda) = \frac{2\tau(4\omega)^\nu}{\nu(\omega-1)(\omega+1)^{2\nu}}  \\  \nonumber
&& \times \, _2F_1\biggl(\frac{\nu(1-1/\omega)}{2}, \nu ;\, \frac{\nu(1-1/\omega)}{2}+1;\, \frac{(\omega-1)^2}{(\omega+1)^2}\biggr).
\end{eqnarray}

It is worth noting that many peculiar properties of heterogeneous
diffusion-limited reactions and related first-passage phenomena
originate from the average over the initial diffusivity $D_0$.  For
instance, for a given realization of $D_0$, the short-time behavior of
the probability density $\rho(t|\x_0)$ is determined by
Eq. (\ref{eq:rhoBM_short}) for homogeneous diffusion (with $\bar{D}$
replaced by $D_0$).  This probability density function is very
sensitive to the chosen $D_0$ and exhibits strong variations between
random realizations of $D_0$, particularly at short times.  This
observation naturally raises the question of a reliable interpretation
of single-particle realizations and their ensemble average.  Moreover,
an empirical average over a finite number of realizations depends on
that number and thus may lead to transient regimes.  This is
particularly clear within the superstatistical approximation (see
Sec. \SISS for details) when the exponential function
$\exp(-D_0t\lambda_n)$ from the spectral decomposition of the
propagator for homogeneous diffusion is averaged over $D_0$ drawn from
the stationary Gamma distribution (\ref{eq:Gamma}): while the exact
average (over infinitely many realizations) gives a power law $(1 +
\bar{D} t\lambda_n/\nu)^{-\nu}$, an empirical average over a finite
number of realizations yields a linear combination of exponential
functions and thus, ultimately, decays exponentially.  When the number
of realizations increases, this linear combination becomes closer and
closer to the power law at intermediate times, but this regime is
still terminated by an exponential cut-off.  In other words, the
ensemble average over the initial diffusivity accurately describes the
diffusion-reaction properties of a heterogeneous medium if the number
of realizations is large enough.

\subsection*{2. Superstatistical approximation}

Although we have derived in Sec. \SIgeneral the exact form of the
propagator and related quantities, their short-time behavior is
determined by infinitely many eigenmodes and thus remains challenging
to access.  To overcome this difficulty, one can resort to a
superstatistical approximation \cite{Beck2003,Beck2005}, a common
simplified way for accounting for diffusivity heterogeneities.  In a
nutshell, the effect of disorder is approximately incorporated by
assuming that a particle diffuses with a constant but randomly chosen
diffusivity $D_0$, whereas the resulting propagator and related
quantities are obtained by averaging over the distribution of the
initial diffusivity $D_0$.  As discussed in
\cite{Chechkin2017,Lanoiselee2018}, the superstatistical description
accurately approximates the propagator in $\R$ at short times, $t \ll
\tau$, when the stochastic diffusivity $D_t$ does not evolve too far
from its initial value $D_0$, but fails at long times.  It is
instructive to compare this approximation to our exact solution.  The
propagator for homogeneous diffusion with a constant diffusivity $D_0$
admits a spectral decomposition
\begin{equation}  
P_{\rm hom}(\x,t|\x_0) = \sum\limits_{n=1}^\infty u_n(\x) \, u_n(\x_0) \, \exp(-D_0 t\lambda_n) .
\end{equation}
Since diffusivity heterogeneities in a stationary regime are described
by the Gamma distribution (\ref{eq:Gamma}), the average of the
propagator with this distribution yields
\begin{equation}  \label{eq:P_SS}
P_0(\x,t|\x_0) = \sum\limits_{n=1}^\infty u_n(\x) \, u_n(\x_0) \, \bigl(1 + \lambda_n \bar{D} t/\nu\bigr)^{-\nu} ,
\end{equation}
where the subscript $0$ highlights the short-time range of validity of
this superstatistical approximation.  One also approximates the
probability density function of the first-passage time as
\begin{equation}  \label{eq:rho_SS}
\rho_0(t|\x_0) = \bar{D} \sum\limits_{n=1}^\infty \frac{\lambda_n  \, u_n(\x_0)}{\bigl(1 + \lambda_n \bar{D} t/\nu\bigr)^{\nu+1}} \, 
\int\limits_\Omega d\x \, u_n(\x).
\end{equation}
In particular, the propagator and the probability density function
exhibit a power-law long-time decay that disagrees with the
exponential decay discussed in the Result section.  Nevertheless, we
will show below that these superstatistical approximations are
accurate at short times.

We focus on the short-time behavior of the probability density
function $\rho(t|\x_0)$ of the first-passage time to a perfectly
reactive region $\Gamma$ on the boundary of the confining domain
$\Omega$.  For Brownian motion with diffusivity $D_0$, the short-time
behavior of this density is well known:
\begin{equation}  \label{eq:rho_BM0}
\rho_{\rm hom}(t|\x_0) \simeq \frac{\delta}{\sqrt{4\pi D_0 t^3}} \, \exp\bigl( - \delta^2/(4D_0 t) \bigr),
\end{equation}
where $\delta$ is the distance from the starting point $\x_0$ to the
reactive region $\Gamma$.   As a very fast arrival to the
reactive region is realized by a ``direct trajectory''
\cite{Godec2016} from $\x_0$ to the closest points on $\Gamma$,
the right-hand side of Eq. (\ref{eq:rho_BM0}) is close to the exact
probability density function of the first-passage time to an absorbing
point on the half-line \cite{Redner} (see also Sec. \SIunbounded).
The average of Eq. (\ref{eq:rho_BM0}) with the Gamma distribution
(\ref{eq:Gamma}) yields the short-time behavior of the probability
density function:
\begin{equation}  \label{eq:rho_half_SS}
\rho(t|\x_0) \simeq  \frac{2^{1/2-\nu}}{\Gamma(\nu) \sqrt{\pi} \, t} \, z_0^{\nu+1/2}  \, K_{\nu-1/2}(z_0) ,
\end{equation}
where $z_0 = \delta \sqrt{\nu/(\bar{D}t)}$, and $K_\nu(z)$ is the
modified Bessel function of the second kind.  As $t\to 0$, one has
$z_0\to\infty$, and the asymptotic behavior of $K_\nu(z)$ yields
\begin{equation}  \label{eq:Srho_short}
\rho(t|\x_0) \simeq  t^{-1} \, \frac{2^{-\nu}}{\Gamma(\nu)} \, \biggl(\delta \sqrt{\nu/(\bar{D} t)}\biggr)^{\nu} \, e^{- \delta \sqrt{\nu/(\bar{D}t)}}  \,.
\end{equation}
We note that the numerical prefactor can be affected by the geometric
structure of the domain.  For instance, if the domain is an interval
and the particle starts from the middle, then both absorbing endpoints
are equally accessible that doubles chances to reach the target at
short times, and the asymptotic relation (\ref{eq:Srho_short}) should
be multiplied by $2$.  Ignoring the numerical prefactor, one gets
Eq. (\ref{eq:rho_short}).

The short-time asymptotic relation (\ref{eq:Srho_short}) is valid as
soon as $\delta \sqrt{\nu/(\bar{D}t)} \gg 1$ and $t \ll
\tau$ that can be written as
\begin{equation}
t/\tau \ll \min\{1, ~ \delta^2/(\sigma^2 \tau^2)\} \,.
\end{equation}
When the distance to the target $\delta$ is greater than the disorder
length scale $\sigma\tau$, the accuracy of the short-time relation is
only limited by the time scale $\tau$.  In turn, when $\delta < \sigma
\tau$, the major limitation is $t \ll \nu \delta^2/\bar{D}$.

\begin{figure}[t!]
\begin{center}  
\includegraphics[width=85mm]{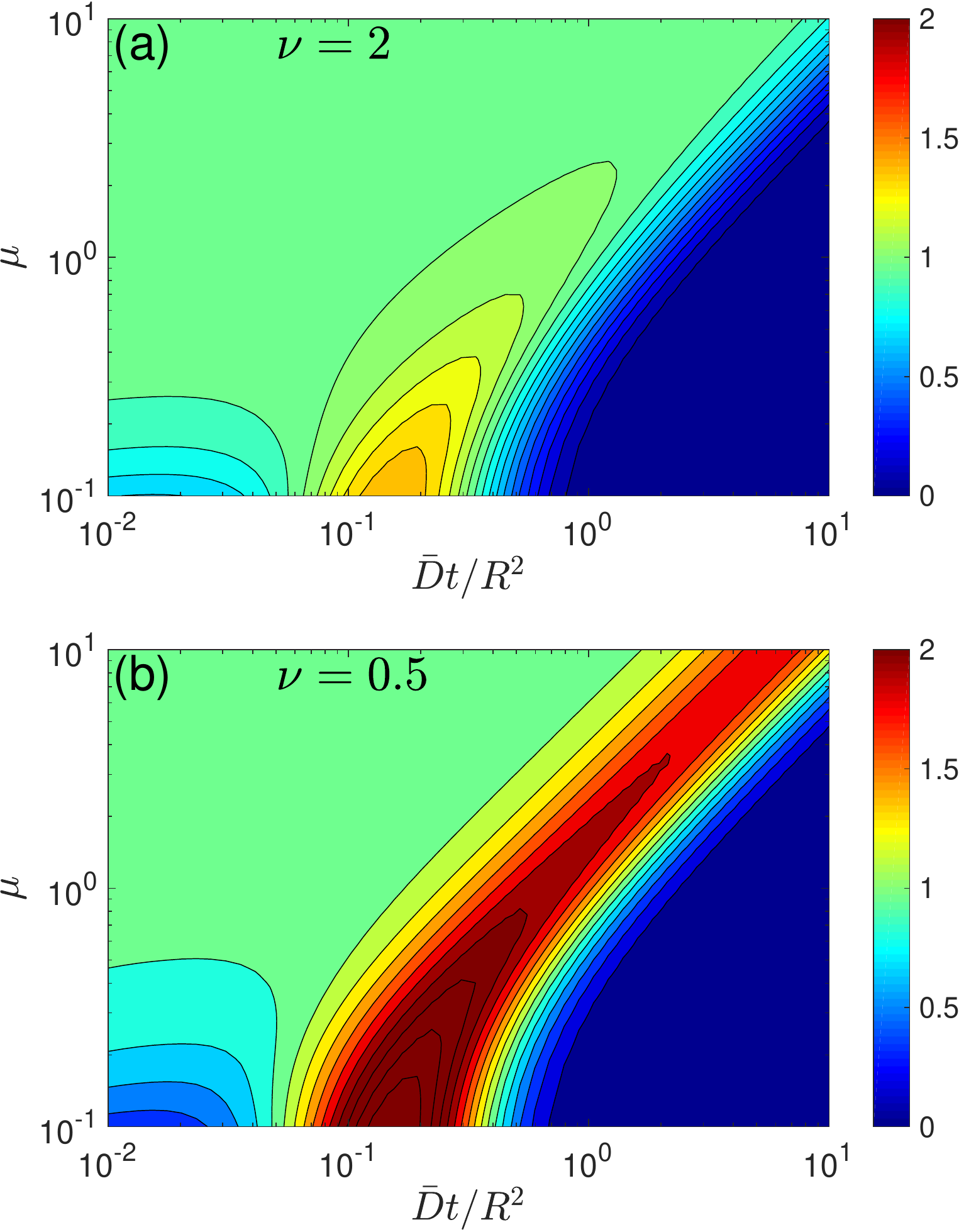} 
\end{center}
\caption{
{\bf The quality of the superstatistical approximation}.  Illustration
for the probability density function of the first-passage time from
the center to the perfectly reactive boundary of a ball of radius $R$.
The ratio between the exact solution $\rho(t|0)$ from
Eq. (\ref{eq:rho_sphere}) and its superstatistical approximation
$\rho_0(t|0)$ from Eq. (\ref{eq:rho_sphere_SS}) is encoded by color
and shown by 16 contour plots, for $\nu = 2$ {\bf (a)} and $\nu = 0.5$
{\bf (b)}.  The approximation is accurate when the ratio is close to
$1$ (left upper corner).  Here $\nu = \bar{D}/(\tau\sigma^2)$ and $\mu
= \sigma\tau/R$.}
\label{fig:error}
\end{figure}

Figure \ref{fig:error} illustrates the quality of the superstatistical
approximation of the probability density $\rho(t|0)$ of the
first-passage time from the center to the perfectly reactive boundary
of a ball of radius $R$.  In this case, the superstatistical
approximation (\ref{eq:rho_SS}) reads
\begin{equation}  \label{eq:rho_sphere_SS}
\rho_0(t|0) = 
\frac{2\pi^2\bar{D}}{R^2} \sum\limits_{n=1}^\infty n^2(-1)^{n-1} 
\biggl(1 + \frac{\bar{D} t\pi^2 n^2}{\nu R^2}\biggr)^{-\nu-1} \, .
\end{equation}
Note that this superstatistical approximation does not depend on the
disorder scale $\mu = \sigma\tau/R$.  When the disorder scale $\mu$ is
large, the diffusivity $D_t$ does not change much from its randomly
chosen starting value $D_0$, so that the superstatistical
approximation is accurate for a broad range of times (left upper
corner of contour plots in Fig. \ref{fig:error}).  In this regime,
deviations appear only at relatively long times (right upper corner).
As the disorder scale $\mu$ decreases, the validity range of the
superstatistical approximation progressively shrinks toward very short
times.  This is true for both weak ($1/\nu = 0.5$) and strong ($1/\nu
= 2$) disorder, deviations being higher in the latter case.  We
conclude that the superstatistical approximation and the resulting
short-time behavior are accurate when $\mu$ is not too small.


\subsection*{3. Limiting behavior of the probability density function}

Figure \ref{fig:rhot_int} illustrated the behavior of the probability
density function $\rho(t|\x_0)$ of the first-passage time to a
perfectly reactive surface of a ball of radius $R$.  We explored the
space $(\mu,1/\nu)$ of parameters characterizing the scale and the
strength of the dynamic disorder, respectively:
\begin{equation}  \label{eq:munu}
\mu = \sigma\tau/R ,   \qquad 1/\nu = \tau \sigma^2/\bar{D}.  
\end{equation}
In particular, we studied the impact of these parameters onto the
short-time and long-time tails of the probability density.  In this
section, we investigate the behavior of the probability density
$\rho(t|\x_0)$ in two limits: $\mu \to 0$ (with fixed $1/\nu$) and
$1/\nu \to 0$ (with fixed $\mu$).

We first recall that the function $\Upsilon(t;\lambda)$ from
Eq. (\ref{eq:Ups}) converges to
\begin{equation}
\Upsilon_{\rm hom}(t;\lambda) = \exp(-\bar{D}t\lambda)
\end{equation}
for homogeneous diffusion when either the amplitude $\sigma$ of
diffusivity fluctuations vanishes (with fixed $\tau$), or their time
scale $\tau$ vanishes (with fixed $\sigma$).  However, the limiting
behavior for $\mu\to 0$ or $1/\nu \to 0$ is more intricate.  Fixing
$R$ and $\bar{D}$, one can express $\sigma$ and $\tau$ from
Eq. (\ref{eq:munu}) as
\begin{equation}
\sigma = \frac{\bar{D}}{R} \, \frac{1}{\mu \nu} \,,  \qquad \tau = \frac{R^2}{\bar{D}} \, \nu \mu^2 .
\end{equation}
The limit $\mu \to 0$ (with fixed $1/\nu$) implies the double limit
$\sigma\to\infty$ and $\tau \to 0$, i.e., fluctuations of diffusivity
become giant but rapidly reverting to the mean.  In turn, the limit
$1/\nu \to 0$ (with fixed $\mu$) implies the double limit $\sigma\to
0$ and $\tau \to \infty$, i.e., fluctuations of diffusivity are small
but very slowly reverting to the mean.  It is thus not clear, {\it a
priori}, whether a diffusing particle would manage to average out such
diffusivities to be described by homogeneous diffusion.

To clarify these points, we rewrite the function $\Upsilon(t;\lambda)$
in terms of $\nu$ and $\mu$ as
\begin{equation}  \label{eq:Ups_auxil}
\Upsilon(t;\lambda) = \biggl(\frac{4\omega\, e^{-(\omega-1) \hat{t}/(2\nu \mu^2)}}{(\omega+1)^2 
- (\omega-1)^2 e^{- \omega \hat{t}/(\nu \mu^2)}}\biggr)^\nu  \,, 
\end{equation}
where $\hat{t} = \bar{D} t/R^2$ is the rescaled time, and $\omega =
\sqrt{1 + 4\mu^2 R^2\lambda}$.  

In the limit $1/\nu \to 0$ (with fixed $\mu$), one can expand the
exponential function $e^{- \omega \hat{t}/(\nu \mu^2)}$ in the
denominator of Eq. (\ref{eq:Ups_auxil}) to get, for a fixed $t$,
\begin{equation*}
\Upsilon(t;\lambda) \simeq \Upsilon_{\rm hom}(t;\lambda) + O(1/\nu).
\end{equation*}
As a consequence, the probability density $\rho(t|\x_0)$ approaches
that for homogeneous diffusion as $1/\nu$ is getting smaller.  Since
the time $t$ stands in the small expansion parameter, the functions
$\rho(t|\x_0)$ and $\rho_{\rm hom}(t|\x_0)$ are closer to each other
for smaller $t$.

In the other limit $\mu \to 0$ (with $1/\nu$ fixed), one uses the
Taylor expansion $\omega \simeq 1 + 2\mu^2 R^2\lambda + O(\mu^4)$ for
a fixed $\lambda$ to show that
\begin{equation}  \label{eq:Ups_mu0}
\Upsilon(t;\lambda) \simeq \Upsilon_{\rm hom}(t;\lambda) + O(\mu^2).
\end{equation}
One sees again that the function $\Upsilon(t;\lambda)$ converges to
that for homogeneous diffusion in this limit.  However, the
probability density $\rho(t|\x_0)$ remains different from $\rho_{\rm
hom}(t|\x_0)$ at short times.  In fact, the expansion
(\ref{eq:Ups_mu0}) holds for any fixed $\lambda$, whereas the spectral
decomposition (\ref{eq:rho_main}) involves terms with Laplacian
eigenvalues $\lambda_n$ that grow to infinity as $n$ increases (here
we assume that $\lambda_n$ are enumerated in an increasing order).
Regardless of the smallness of the parameter $\mu > 0$, there exists
an index $n_0$ such that $\mu^2 R^2 \lambda_n \gg 1$ for all $n > n_0$
so that the above expansion is not applicable.  In other words, for
any $\mu > 0$, there remain infinitely many terms
$\Upsilon'(t;\lambda_n)$ that significantly differ from
$\Upsilon'_{\rm hom}(t;\lambda_n)$.  As these terms determine the
short-time asymptotic behavior of the probability density function,
$\rho(t|\x_0)$ exhibits deviations from $\rho_{\rm hom}(t|\x_0)$ at
(very) short times for any $\mu > 0$.

\subsection*{4. Unbounded domains}

The derivation of the propagator in Sec. \SIgeneral is applicable for
any bounded domain $\Omega\subset \R^d$, for which the eigenvalue
problem for the Laplace operator is well defined, and the spectrum is
known to be discrete.  An extension to unbounded domains should handle
the continuous spectrum of the Laplace operator, in particular, the
absence of $L^2$-normalized eigenfunctions.  For instance, the
propagator for the whole line $\R$ derived in
\cite{Lanoiselee2018} admits a form similar to Eq. (\ref{eq:main}), in
which the eigenvalues $\lambda_n$ are replaced by $q^2$, the
eigenfunctions $u_n(x)$ and $u_n(x_0)$ are replaced by Fourier modes
$e^{iq x}$ and $e^{-iqx_0}$, and the sum is turned into the integral
over $q$:
\begin{equation}
P_{\R}(x,t|x_0) = \int\limits_{-\infty}^\infty \frac{dq}{2\pi} \, e^{iq(x-x_0)} \, \Upsilon(t;q^2).
\end{equation}
To illustrate the impact of dynamic disorder in the case of unbounded
domains, we focus on two important examples: the half-space and the
exterior of a ball.  For these examples, one can use the known form of
the propagator in $\R^d$ and apply the image method.

\subsubsection*{Half-space}

The propagator in $\R^d$ was derived in \cite{Lanoiselee2018} in the
form
\begin{equation}  \label{eq:Pd}
P_{\R^d}(\x,t|\x_0) = \int\limits_{\R^d} \frac{d\q}{(2\pi)^d} \, e^{i\q(\x-\x_0)} \, \Upsilon(t;|\q|^2).
\end{equation}
In contrast to the Gaussian propagator for homogeneous diffusion, the
propagator $P_{\R^d}(\x,t|\x_0)$ in $d$ dimensions is not the product
of $d$ one-dimensional propagators because $\Upsilon$ is not an
exponential function of $|\q|^2$.  This is expected because the
motions along different directions are correlated via the stochastic
diffusivity $D_t$.

The propagator in a half-space $\R^d_+$ with a perfectly reactive
hyperplane can be obtained by the image method:
\begin{equation}
P_{\R^d_+}(\x,t| \x_0) = P_{\R^d}(\x,t| \x_0) - P_{\R^d}(\x,t| \hat{\x}_0)  ,
\end{equation}
where $\hat{\x}_0$ is the mirror reflection of $\x_0$ with respect to
the reactive hyperplane.  The survival probability in the half-space
is deduced by integrating this propagator over $\x \in \R^d_+$.
Importantly, the statistics of the first-passage time to the reactive
hyperplane is not affected by the lateral motion (that is parallel to
the reactive hyperplane), as for homogeneous diffusion.  In fact, the
integral of the propagator $P_{\R^d}(\x,t|\x_0)$ over all lateral
coordinates yields the one-dimensional propagator in the orthogonal
direction (that we choose to be along $x_1$ for clarity):
\begin{equation}
\int\limits_{\R^{d-1}} dx_2 \ldots dx_d \, P_{\R^d}(\x,t|\x_0) = P_\R(x_1,t|x_{0,1}).
\end{equation}
In other words, the computation of the survival probability and the
probability density function of the first-passage time in the
half-space is reduced to that for a half-line with an absorbing
endpoint.  We focus thus on this one-dimensional problem.

Using the image method, we deduce the propagator on the half-line
$(0,\infty)$ with an absorbing endpoint at $0$:
\begin{align}  \nonumber
P(x,t| x_0) &= P_{\R}(x,t| x_0) - P_{\R}(x,t| -x_0) \\
&= -\frac{i}{\pi} \int\limits_{-\infty}^{\infty} dq \, e^{iqx} \, \sin(qx_0)\, \Upsilon(t;q^2) .
\end{align}
This is the probability density for a particle started at $x_0 > 0$ to
be at $x \geq 0$ at time $t$, without hitting the absorbing endpoint
$0$ on its way.  Integrating the propagator over the arrival point
$x$, one gets the survival probability $S(t| x_0)$
\begin{equation}
S(t| x_0)=\int\limits_0^\infty dx \, P(x,t\vert x_0) = \frac{2}{\pi} \int\limits_0^{\infty} \frac{dq}{q} \, \sin(qx_0) \, \Upsilon(t;q^2),
\end{equation}
from which the probability density function of the first-passage time
is
\begin{equation}  \label{eq:rho_half}
\rho(t| x_0) = -\frac{\partial S(t| x_0)}{\partial t} = - \frac{2}{\pi} \int\limits_0^{\infty}\frac{dq}{q}\, \sin(qx_0) \, 
\Upsilon'(t;q^2) ,
\end{equation}
where $\Upsilon'(t;\lambda)$ is given by Eq. (\ref{eq:dP}).  For
comparison, the probability density function for homogeneous diffusion
with diffusivity $\bar{D}$ is
\begin{equation} \label{eq:rho_BM}
\rho_{\rm hom}(t|x_0) = \frac{x_0}{\sqrt{4\pi \bar{D} t^3}} \exp\left(-\frac{x_0^2}{4\bar{D} t}\right).
\end{equation}

In the long-time limit, the terms $e^{-\omega t/\tau}$ in
Eq. (\ref{eq:dP}) vanish, yielding
\begin{equation}
\rho(t| x_0)\simeq \hspace*{-0.5mm} \int\limits_0^{\infty}\frac{\nu dq}{\pi \tau q}\sin(qx_0)
(\omega-1) e^{-\frac{\nu}{2}\left(\omega-1\right) \frac{t}{\tau}} \biggl(\frac{4\omega}{\left(\omega+1\right)^2} \biggr)^\nu ,
\end{equation}
with $\omega = \sqrt{1 + 4\sigma^2\tau^2 q^2}$.  Changing the
integration variable and eliminating all terms of order $1/t$ or
higher, one gets the classic power-law behavior
\begin{equation}
\rho(t| x_0) \simeq \frac{x_0}{\sqrt{4\pi \bar{D} t^3}} \qquad (t\to\infty),
\end{equation}
that corresponds to Brownian motion (cf. Eq. (\ref{eq:rho_BM})).  In
this limit, the particle has enough time to average out the disorder
in diffusivities and thus behaves as a Brownian particle with the mean
diffusivity $\bar{D}$.  This conclusion contrasts with the case of
bounded domains, for which the long-time asymptotic behavior could be
significantly affected by the disorder (see Eq. (\ref{eq:rho_long})
and the related discussion in the Results section).  The main
difference for unbounded domains is the absence of the largest
geometric length scale (an analog of $\lambda_1^{-1/2}$) as the
Laplacian spectrum is continuous and bounded by zero.  From a
practical point of view, the long-time behavior is dominated by very
long trajectories exploring the unbounded domain so that diffusivity
heterogeneities are averaged out independently of their length scale
$\sigma\tau$.  In particular, the mean FPT is infinite, as for
homogeneous diffusion.

The short-time behavior can be obtained via the superstatistical
approach by averaging the Gaussian propagator for Brownian motion with
the Gamma distribution (\ref{eq:Gamma}) for diffusivities and then
applying the image method.  First, one gets the averaged propagator in
$\R$
\begin{equation}
P_{\R,0}(x,t|x_0) = \frac{\sqrt{\nu}}{\sqrt{\bar{D} t}} \,  \K_\nu\biggl(|x-x_0| \sqrt{\nu/(\bar{D} t)}\biggr) ,
\end{equation}
where we defined
\begin{equation}  \label{eq:Knu}
\K_\nu(z) = \frac{2^{1/2-\nu}}{\Gamma(\nu) \sqrt{\pi}} \, z^{\nu-1/2} \, K_{\nu-\frac12}(z) .
\end{equation}
The image method yields the averaged propagator on the half-line, from
which the superstatistical approximation of the survival probability
follows
\begin{equation} 
S_0(t|x_0) = 2 \int\limits_0^{z_0} dz \, \K_\nu(z),
\end{equation}
where $z_0 = x_0 \sqrt{\nu/(\bar{D} t)}$.  This integral can be
expressed via Struve functions.  In turn, the superstatistical
approximation of the probability density function is much simpler:
\begin{equation}  \label{eq:rho_half_SS2}
\rho_0(t|x_0) = \frac{z_0 \, \K_\nu(z_0)}{t} \, .
\end{equation}
Naturally, we retrieved the right-hand side of
Eq. (\ref{eq:rho_half_SS}), which was obtained as an approximate
asymptotic relation for bounded domains.  As $t\to 0$, this relation
leads to the asymptotic behavior (\ref{eq:rho_short}).  As expected,
the short-time behavior does not depend on the type of the confining
domain.  It is worth noting that for the half-line, the
superstatistical approach captures qualitatively even the long-time
asymptotic behavior for $\nu > 1/2$:
\begin{equation}  \label{eq:rho_half_SS3}
\rho_0(t|x_0) \simeq \frac{\Gamma(\nu-1/2) \sqrt{\nu}}{\Gamma(\nu)}\,  \frac{x_0}{\sqrt{4\pi\bar{D} t^3}}  \quad (t\to\infty) ,
\end{equation}
but overestimates the probability density by a numerical factor
depending only on $\nu$.

\begin{figure}[t!]
\begin{center}  
\includegraphics[width=85mm]{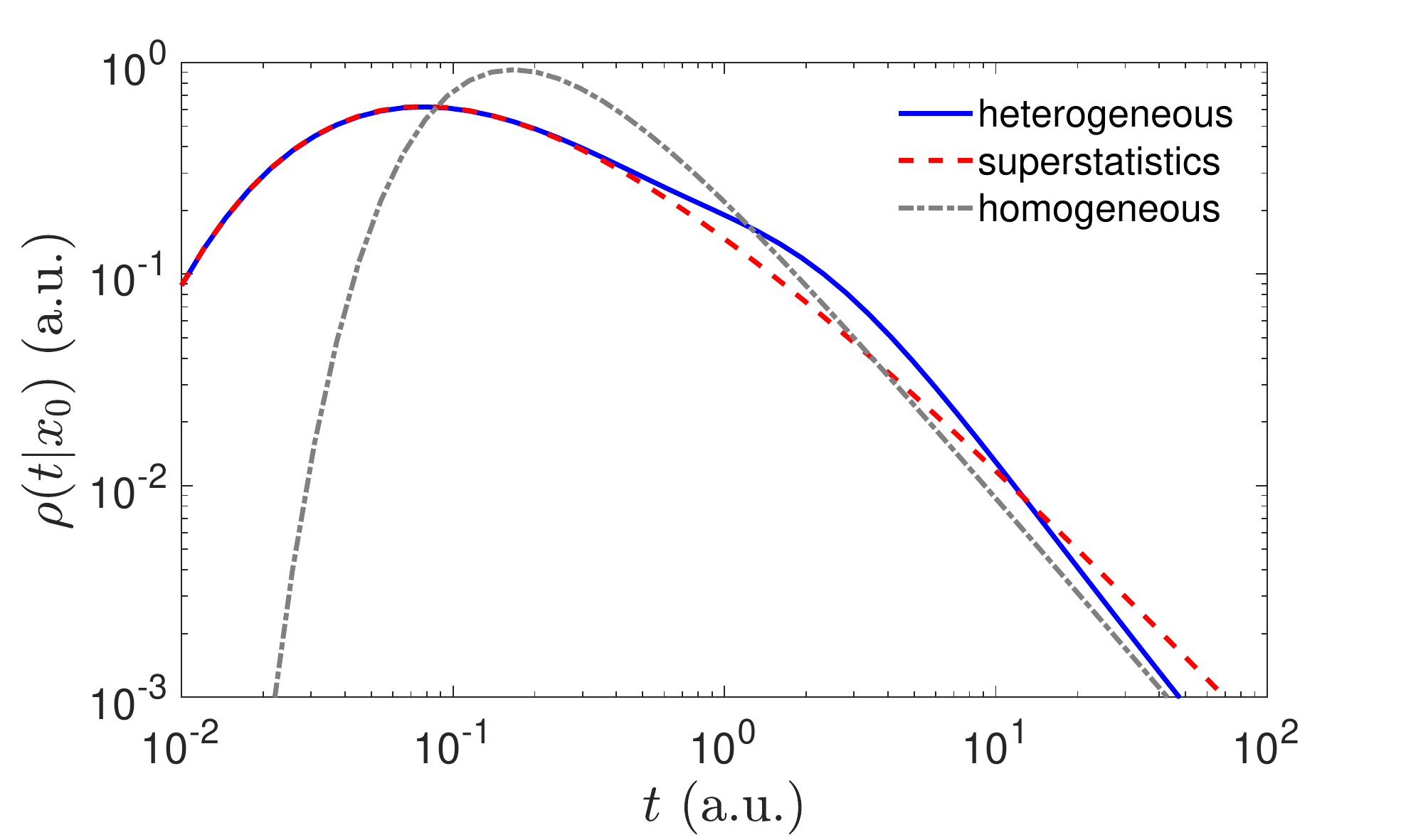}
\end{center}
\caption{
{\bf First-passage times on the half-line.}  Probability density
function $\rho(t|x_0)$ of the first-passage time to the absorbing
endpoint $0$ of the half-line $(0,\infty)$, with $\bar{D} = 1$,
$\sigma = 1$, $x_0 = 1$, and $\tau = 2$ (here arbitrary units are
used).  The exact solution (\ref{eq:rho_half}) (solid line) is
compared to the superstatistical approximation (\ref{eq:rho_half_SS2})
(dashed line) and the probability density function (\ref{eq:rho_BM})
for homogeneous diffusion (dash-dotted line).}
\label{fig:rhot_half}
\end{figure}

Figure \ref{fig:rhot_half} compares the exact probability density
$\rho(t|x_0)$ from Eq. (\ref{eq:rho_half}), its superstatistical
approximation (\ref{eq:rho_half_SS2}), and the probability density
(\ref{eq:rho_BM}) for homogeneous diffusion.  One can see that the
superstatistical approximation turns out to be very accurate not only
at short times, but also at intermediate times.  At long times, this
approximation provides the correct power law $t^{-3/2}$ but
overestimates the prefactor (cf. Eq. (\ref{eq:rho_half_SS3})).  In
turn, the probability density function for Brownian motion yields the
correct long-time asymptotic behavior but generally fails.


As discussed in the Results section, the macroscopic reaction rate
$J(t)$ (i.e., the diffusive flux onto the reactive target) can be
obtained by averaging the probability density $\rho(t|x_0)$ with a
prescribed initial concentration of particles $c_0(x_0)$:
\begin{equation}
J(t) = \int\limits_0^\infty dx_0 \, c_0(x_0) \, \rho(t|x_0) .
\end{equation}
Setting a uniform initial concentration $c_0$ and using the
probability density function in Eq. (\ref{eq:rho_half}), we get
\begin{equation}
J(t) = - \frac{2c_0}{\pi} \int\limits_0^\infty dx_0 \int\limits_0^\infty dq \, \frac{\sin (qx_0)}{q} \, \Upsilon'(t;q^2) .
\end{equation}
To evaluate the integral, we introduce an auxiliary integral
\begin{eqnarray}  \nonumber
I_s(t) &=& \int\limits_0^\infty dx_0 \, e^{-sx_0} \int\limits_0^\infty dq \, \frac{\sin qx_0}{q} \, \Upsilon'(t;q^2)  \\
\label{eq:Q_auxil2}
&=& \int\limits_0^\infty \frac{dq \,\Upsilon'(t;q^2)}{s^2 + q^2} 
\end{eqnarray}
and then get the macroscopic reaction rate
\begin{equation}
J(t) = - \frac{2c_0}{\pi} \lim\limits_{s\to 0} I_s(t) 
= -  \frac{2c_0}{\pi} \int\limits_0^\infty \frac{dq \, \Upsilon'(t;q^2)}{q^2} \,.
\end{equation}
As expected for one-dimensional setting, the macroscopic reaction rate
vanishes in the long-time limit as all diffusing particles are
progressively absorbed and finally exhausted.

\subsubsection*{Exterior of a ball}

We provide the exact solution to another important example of an
unbounded domain -- the exterior of a ball of radius $R$ with
perfectly reactive boundary.  Since the seminal work by Smoluchowski
\cite{Smoluchowski1917}, this is an emblematic problem of
diffusion-limited reactions.

\paragraph*{Survival probability.}

It is convenient to use the representation of the propagator
(\ref{eq:Pd}) in spherical coordinates derived in
\cite{Lanoiselee2018}:
\begin{equation}  \label{eq:P_Rd}
P_{\R^d}(\x,t|\x_0) = \frac{\delta^{1-d/2}}{(2\pi)^{d/2}}\int\limits_0^\infty dq\, q^{d/2} \, J_{\frac{d}{2}-1}(q\delta) \, \Upsilon(t; q^2),
\end{equation}
where $\delta = \|\x - \x_0\|$ is the distance between the points $\x$
and $\x_0$, and $J_\nu(z)$ is the Bessel function of the first kind.

In three dimensions ($d = 3$), representing the points $\x$ and $\x_0$
in spherical coordinates with respect to a fixed center and averaging
over the angular coordinates, one can rewrite the above propagator as
\begin{equation}
P_{\R^3}(r,t|r_0) = \hspace*{-1mm} \int\limits_0^\infty \hspace*{-1mm} dq \frac{\cos(q(r-r_0)) - \cos(q(r+r_0))}{4\pi^2 rr_0}  \Upsilon(t; q^2)
\end{equation}
(here we oriented the spherical coordinates in the direction to the
point $\x_0$ so that $\x_0 = (r_0,0,0)$ and used thus $\delta =
\sqrt{r^2 - 2rr_0 \cos\theta + r_0^2}$).  Examining this particular
form, we realize that the propagator outside a ball of radius $R$ with
Dirichlet boundary condition reads
\begin{eqnarray}    \nonumber
P(r,t|r_0) &=& \int\limits_0^\infty dq \, \frac{\cos(q(r-r_0)) - \cos(q(r+r_0-2R))}{4\pi^2 rr_0} \\    \label{eq:P_3D_ext}
           && \quad \times \Upsilon(t;q^2).
\end{eqnarray}
In order to compute the integral over the volume, we first evaluate an
auxiliary integral
\begin{eqnarray} 
I_s(t|r_0) &=& 4\pi \int\limits_R^\infty dr \, r \, e^{-sr} \, P(r,t|r_0) \\  \nonumber
&=& \frac{2e^{-sR}}{\pi r_0} \int\limits_0^\infty dq\, \frac{q \sin(q(r_0-R))}{q^2 + s^2} \, \Upsilon(t;q^2) .
\end{eqnarray}
The derivative of this expression with respect to $s$, evaluated at $s
= 0$ and taken with the sign minus, yields the integral of
$P(r,t|r_0)$ over the volume and thus the survival probability:
\begin{equation}  \label{eq:S_3D_ext}
S(t|r_0) = \frac{r_0 - R}{r_0} + \frac{2R}{\pi r_0} \int\limits_0^\infty dq\, \frac{\sin(q(r_0-R))}{q} \, \Upsilon(t;q^2) . 
\end{equation}
Note that the first term, independent of the function $\Upsilon$,
comes from an accurate evaluation of the limit $s\to 0$ of the
integral term with $s/(q^2 + s^2)^2$ in $\partial I_s(t|r_0)/\partial
s$.  This term is the probability of escaping to infinity.  Note also
that $\Upsilon(t=0;q^2) = 1$ implies the correct initial condition
$S(t=0|r_0) = 1$.  The time derivative of Eq. (\ref{eq:S_3D_ext})
yields
\begin{equation}  \label{eq:rho_3D_ext}
\rho(t|r_0) = - \frac{2R}{\pi r_0} \int\limits_0^\infty dq\, \frac{\sin(q(r_0-R))}{q} \, \Upsilon'(t;q^2) ,
\end{equation}
with $\Upsilon'(t;\lambda)$ given by Eq. (\ref{eq:dP}).  As expected,
this probability density function is not normalized to $1$ because the
probability of escape to infinity is not zero.

The long-time asymptotic relation
\begin{equation}
\Upsilon(t; q^2) \simeq \biggl(\frac{4\omega}{(\omega+1)^2}\biggr)^\nu \exp\biggl(-\frac{2\bar{D} tq^2}{1+\omega}\biggr) 
\end{equation}
implies that the dominant contribution to the integral in
Eqs. (\ref{eq:P_3D_ext}, \ref{eq:S_3D_ext}) comes from $q \approx 0$,
at which $\omega \approx 1$, and thus one gets $\Upsilon(t; q^2)
\approx \exp(-\bar{D} t q^2)$.  As a consequence, the long-time
asymptotic behavior of the propagator, of the survival probability,
and of the probability density function are close to that for Brownian
motion with the constant diffusivity $D = \bar{D}$:
\begin{equation}  \label{eq:3D_prop_BM}
P_{\rm hom}(r,t|r_0) = \frac{\exp\bigl(-\frac{(r-r_0)^2}{4Dt}\bigr) -
\exp\bigl(-\frac{(r+r_0-2R)^2}{4Dt}\bigr)}{8\pi rr_0 \sqrt{\pi Dt}} \,,
\end{equation}
\begin{equation}
S_{\rm hom}(t|r_0) = \frac{r_0 - R}{r_0} + \frac{R}{r_0}\, \erf\bigl((r_0-R)/\sqrt{4Dt}\bigr) , 
\end{equation}
and
\begin{equation} \label{eq:rhot_sphere_ext}
\rho_{\rm hom}(t|r_0) = \frac{R}{r_0} \, \frac{(r_0-R) \exp\bigl(-\frac{(r_0-R)^2}{4Dt}\bigr)}{\sqrt{4\pi Dt^3}}  \,,
\end{equation}
where $\erf(z)$ is the error function.  As for the half-line, the
diffusing particle has enough time to average out heterogeneities of
diffusivities and thus to move asymptotically as via homogeneous
diffusion.

Applying the superstatistical description to the propagator in
Eq. (\ref{eq:3D_prop_BM}) with the Gamma distribution (\ref{eq:Gamma})
for $D$, one finds
\begin{equation}
\begin{split}
P_0(r,t|r_0) & = \frac{1}{4\pi r r_0} \, \frac{\sqrt{\nu}}{\sqrt{\bar{D} t}} 
\biggl\{ \K_\nu \biggl(|r-r_0|\sqrt{\nu/(\bar{D}t)}\biggr)  \\
& - \K_\nu\biggl(|r+r_0-2R|\sqrt{\nu/(\bar{D}t)}\biggr) \biggr\}, \\
\end{split}
\end{equation}
with $\K_\nu(z)$ defined by Eq. (\ref{eq:Knu}).  We get thus the
superstatistical approximation of the survival probability
\begin{equation}
S_0(t|r_0) = 1 - \frac{2R}{r_0}  \int\limits_{z_0}^\infty dz \, \K_{\nu}(z) ,
\end{equation}
with $z_0 = (r_0 - R)\sqrt{\nu/(\bar{D}t)}$, and that of the
probability density function:
\begin{equation}  \label{eq:rhot_sphere_SS}
\rho_0(t|r_0) = \frac{R}{r_0} \, \frac{z_0\, \K_{\nu}(z_0)}{t} \, .
\end{equation}
This superstatistical expression provides the short-time asymptotic
behavior of the exact probability density function.  As expected, this
asymptotic relation is almost identical to its one-dimensional
counterpart in Eq. (\ref{eq:rho_half_SS}), apart from the additional
factor $R/r_0$ accounting for the probability to reach the target.

\begin{figure*}
\begin{center}
\includegraphics[width=\textwidth]{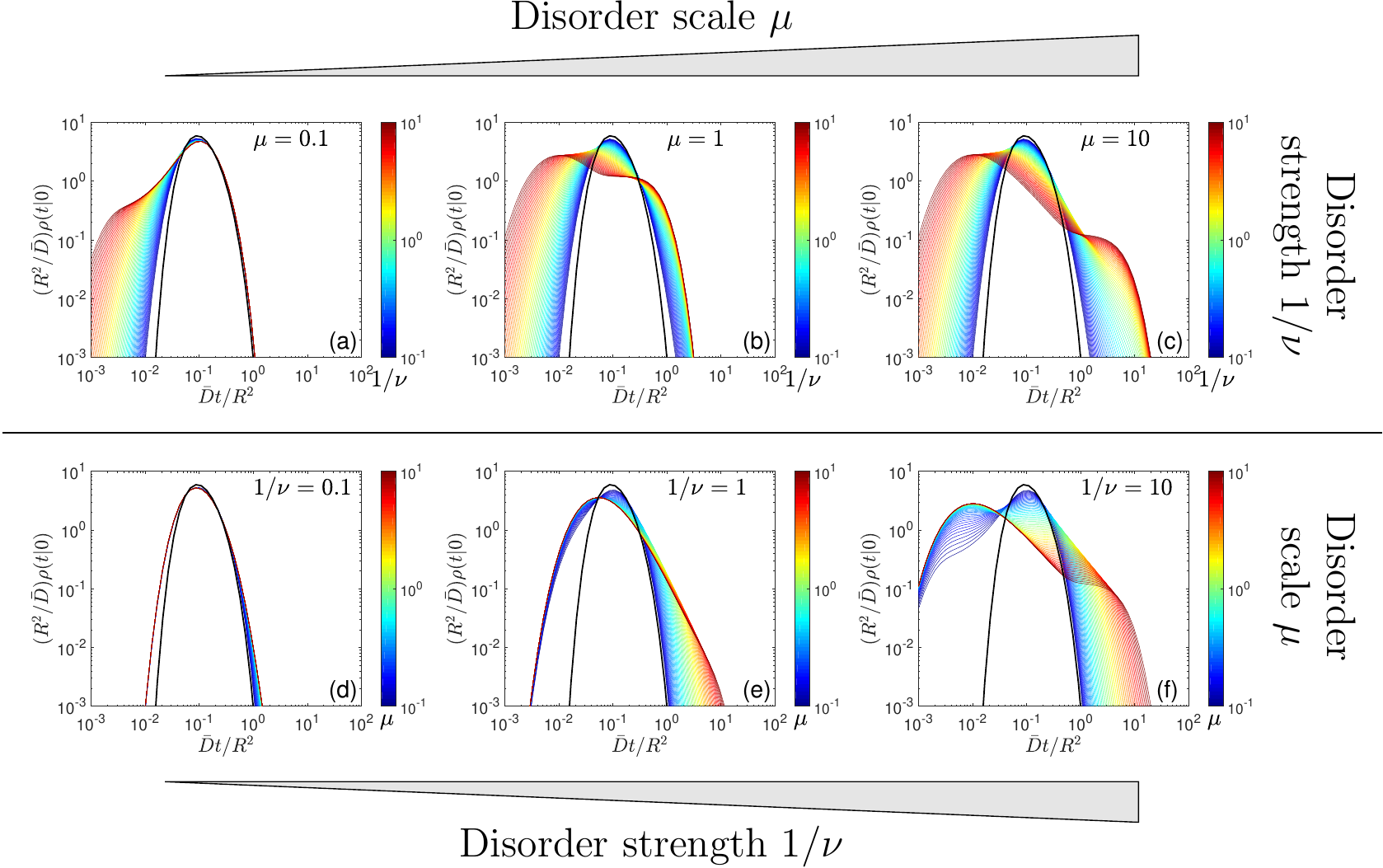}
\end{center}
\caption{
{\bf The impact of dynamic disorder onto the distribution of
first-passage times.}  The probability density function $\rho(t|0)$ of
the first-passage time from the center to the perfectly reactive
boundary of a ball of radius $R$ is shown by colored curves for
various combinations of dimensionless parameters $(\mu,1/\nu)$
characterizing the disorder scale and strength: $\mu = \sigma\tau/R$
and $1/\nu = \tau \sigma^2/\bar{D}$.  Thick black curve presents
$\rho_{\rm hom}(t|0)$ for homogeneous diffusion with diffusivity
$\bar{D}$.  {\bf (a,b,c)}: Three values of the disorder scale $\mu$
($0.1$ {\bf (a)}; $1$ {\bf (b)}; and $10$ {\bf (c)}) and 64 values of
the disorder strength $1/\nu$ in the logarithmic range between
$10^{-1}$ and $10^1$.  {\bf (d,e,f)}: Three values of the disorder
strength $1/\nu$ ($0.1$ {\bf (d)}; $1$ {\bf (e)}; and $10$ {\bf (f)})
and 64 values of the disorder scale $\mu$ in the logarithmic range
between $10^{-1}$ and $10^1$.  Curves encoded by color, ranging from
dark blue ($10^{-1}$) to dark red ($10^1$), as shown by colorbar. }
\label{fig:rhot_int}
\end{figure*}

\begin{figure*}
\begin{center}
\includegraphics[width=\textwidth]{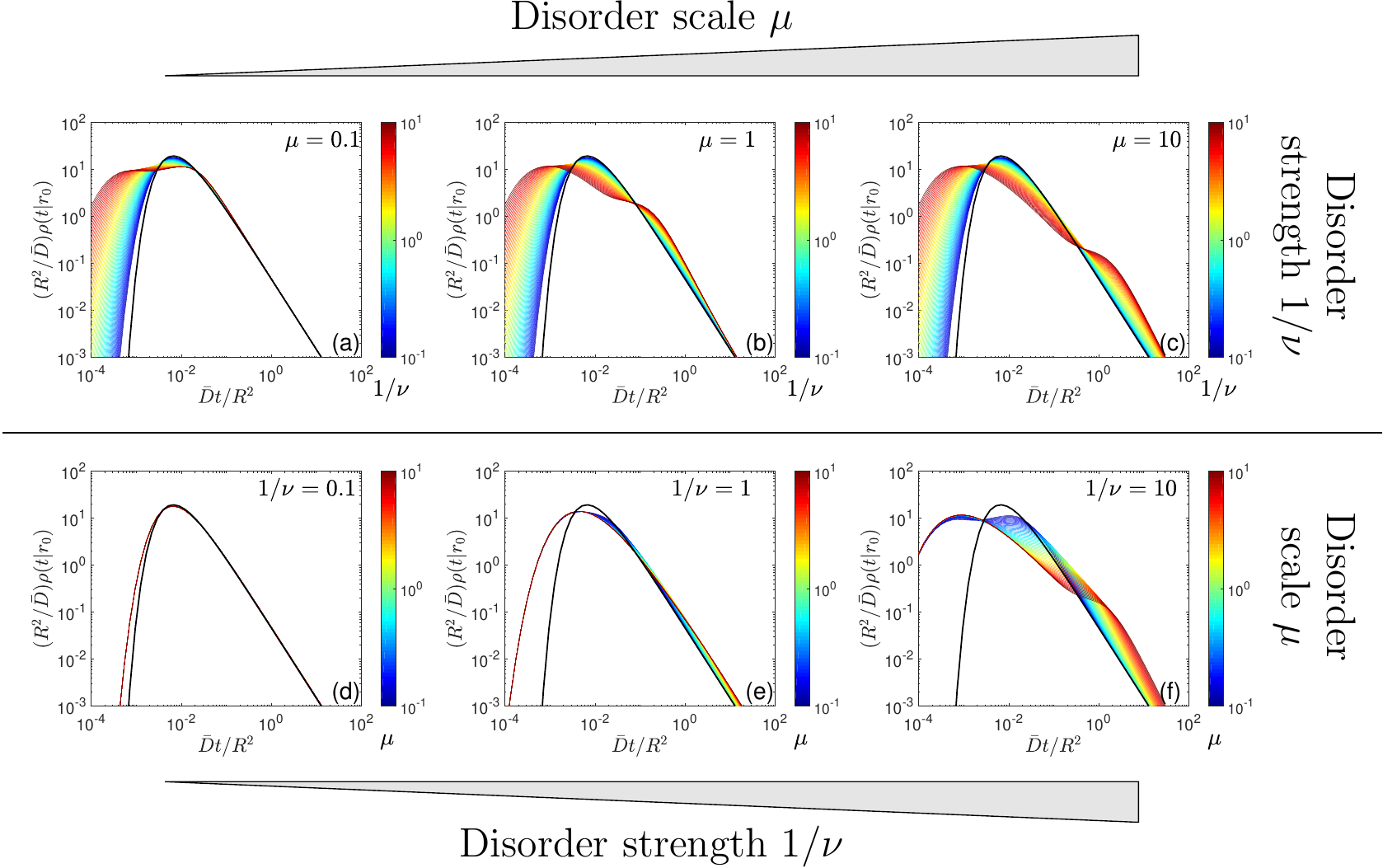}  
\end{center}
\caption{
{\bf The impact of dynamic disorder onto the distribution of
first-passage times for the exterior of a ball.}  The probability
density function $\rho(t|r_0)$ of the first-passage time to a
perfectly reactive ball of radius $R$ from the exterior space (with
$r_0/R = 1.2$) is shown by color lines for various combinations of
dimensionless parameters $(\mu,1/\nu)$ characterizing the disorder
scale and strength: $\mu = \sigma\tau/R$ and $1/\nu = \tau
\sigma^2/\bar{D}$.  Thick black line presents the probability density
$\rho_{\rm hom}(t|r_0)$ from Eq. (\ref{eq:rhot_sphere_ext}) for
homogeneous diffusion with diffusivity $\bar{D}$.  {\bf (a,b,c)}:
Three values of the disorder scale $\mu$ ($0.1$ {\bf (a)}; $1$ {\bf
(b)}; and $10$ {\bf (c)}) and 64 values of the disorder strength
$1/\nu$ in the logarithmic range between $10^{-1}$ and $10^1$.  {\bf
(d,e,f)}: Three values of the disorder strength $1/\nu$ ($0.1$ {\bf
(d)}; $1$ {\bf (e)}; and $10$ {\bf (f)}) and 64 values of the disorder
scale $\mu$ in the logarithmic range between $10^{-1}$ and $10^1$.
Curves encoded by color, ranging from dark blue ($10^{-1}$) to dark
red ($10^1$), as shown by colorbar.}
\label{fig:rho_Sext}
\end{figure*}

Figure \ref{fig:rho_Sext} illustrates the behavior of the probability
density function $\rho(t|r_0)$.  As for Fig. \ref{fig:rhot_int}, we
explore various combinations of dimensionless parameters $(\mu,1/\nu)$
characterizing the disorder scale and strength, in two complementary
ways.  In the top panels (a,b,c), we fix three values of the scale
$\mu$ and range ``continuously'' $1/\nu$ from $10^{-1}$ and $10^1$.
The short-time behavior of the density $\rho(t|r_0)$ (the left tail)
is almost not affected by the scale $\mu$, as expected from the
asymptotic relation (\ref{eq:rho_short}) and the superstatistical
approximation (\ref{eq:rhot_sphere_SS}).  In turn, the long-time
behavior is mostly determined by $\mu$ but also weakly depends on
$\nu$.  For the short-range disorder ($\mu = 0.1$), the right tail
almost coincides with Eq. (\ref{eq:rhot_sphere_ext}) for homogeneous
diffusion, regardless the value of $1/\nu$ in the considered range.
As the scale $\mu$ increases, the particle needs more time to
homogenize stochastic diffusivities, and one observes deviations from
Eq. (\ref{eq:rhot_sphere_ext}), which are larger for stronger disorder
(larger $1/\nu$).  In the bottom panels (d,e,f), we fix three values
of the disorder strength $1/\nu$ and change the scale $\mu$
``continuously''.  One sees again that the left tail is almost
independent of $\mu$, while the right tail exhibits such a dependence.
We stress that variations of this probability density function are in
general lower than that shown in Fig. \ref{fig:rhot_int} for a bounded
domain.  Once again, the exploration of an unbounded domain offers
more opportunities for a diffusing particle to homogenize stochastic
diffusivities at long times.


\paragraph*{Macroscopic reaction rate.}

The macroscopic reaction rate $J(t)$ is obtained by averaging the
probability density function in Eq. (\ref{eq:rho_3D_ext}) with a
uniform initial concentration $c_0$
\begin{equation}
J(t) = - 8c_0 R \int\limits_R^\infty dr_0 \, r_0 \int\limits_0^\infty dq \frac{\sin (q(r_0-R))}{q} \, \Upsilon'(t;q^2) .
\end{equation}
Using again the auxiliary integral (\ref{eq:Q_auxil2}), we get
\begin{equation}
J(t) = - 8c_0 R \lim\limits_{s\to 0} \int\limits_0^\infty \frac{dq \, \Upsilon'(t;q^2)}{s^2 + q^2} \biggl(R + \frac{2s}{s^2+q^2}\biggr).
\end{equation}
Given that $\Upsilon'(t;q^2) \simeq - \bar{D} q^2 + O(q^4)$ as $q\to
0$, the limit of the first term is obtained by setting $s = 0$.  For
the second term, one can extend the integration to $-\infty$ by
symmetry and integrate by parts to get
\begin{eqnarray*}
J(t) &=& - 8c_0 R \biggl\{R \int\limits_0^\infty \frac{dq \, \Upsilon'(t;q^2)}{q^2} \\
&+&
\lim\limits_{s\to 0} \frac12 \int\limits_0^\infty \frac{dq \, s}{s^2 + q^2}  
\frac{\partial}{\partial q} \biggl(\frac{\Upsilon'(t;q^2)}{q}\biggr) \biggr\}.
\end{eqnarray*}
As $s\to 0$, the ratio $s/(s^2+q^2)$ converges to $\pi \delta(q)$
allowing one to evaluate the integral explicitly and yielding
\begin{equation}  \label{eq:Jt}
J(t) = 4\pi c_0 R \bar{D} \biggl(1 - \frac{2R}{\pi \bar{D}} \int\limits_0^\infty dq \, \frac{\Upsilon'(t;q^2)}{q^2} \biggr).
\end{equation}
For instance, one has $\Upsilon'_{\rm hom}(t;q^2) = - \bar{D} q^2
e^{-\bar{D} t q^2}$ for Brownian motion with diffusivity $\bar{D}$,
and thus retrieves the classic Smoluchowski reaction rate
\cite{Smoluchowski1917}:
\begin{equation}   \label{eq:Jt_BM}
J_{\rm hom}(t) = 4\pi c_0 R \bar{D} \biggl(1 + \frac{R}{\sqrt{\pi \bar{D} t}}\biggr).
\end{equation}
In the long-time limit, the second term in both Eqs. (\ref{eq:Jt},
\ref{eq:Jt_BM}) vanishes, and one recovers the Smoluchowski
steady-state reaction rate: 
\begin{equation}  \label{eq:J_Smol}
J_S = 4\pi c_0 \bar{D} R.  
\end{equation}
In turn, the approach to the steady-state solution differs for
homogeneous and heterogeneous cases.

\begin{figure}[t!]
\begin{center}
\includegraphics[width=85mm]{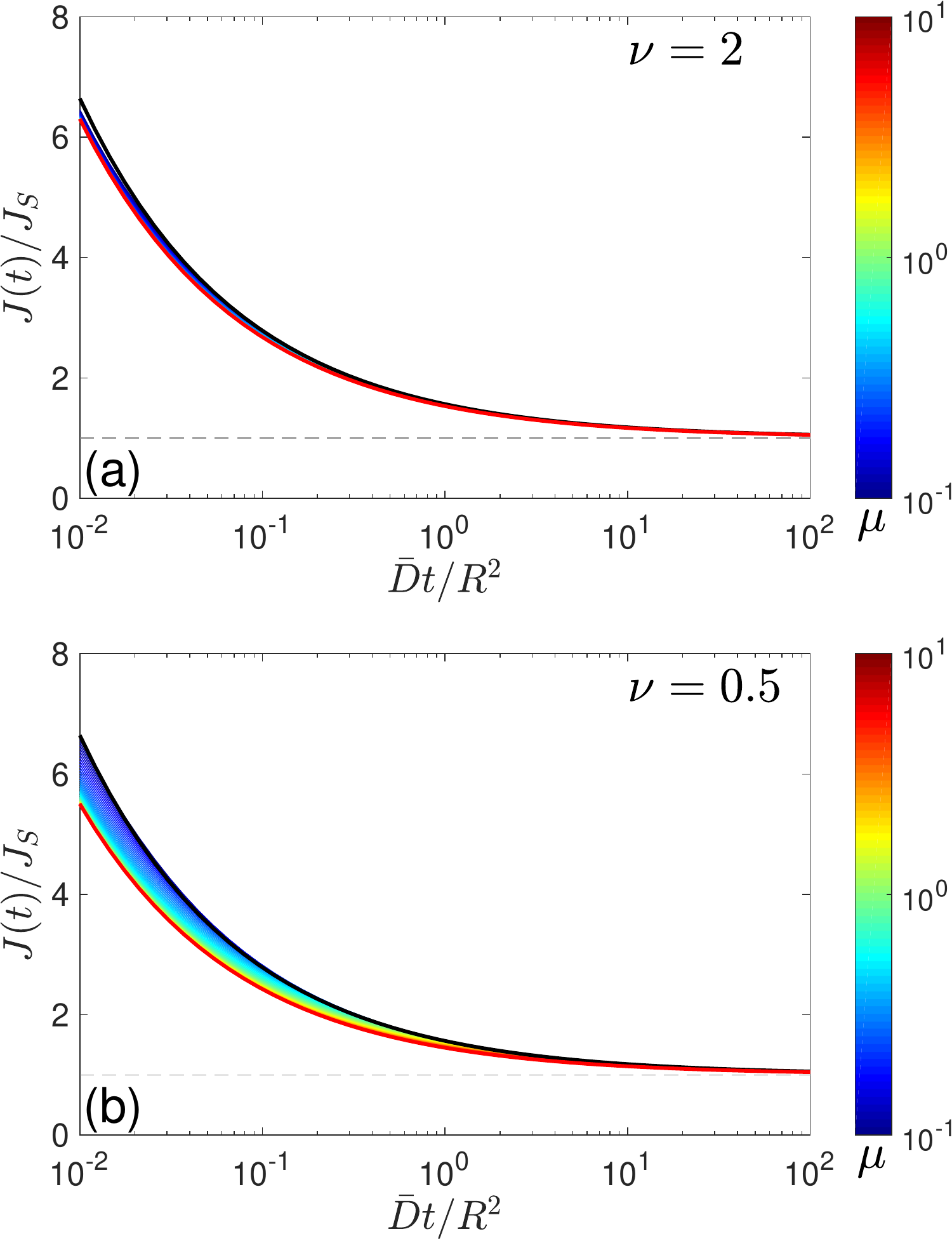}
\end{center}
\caption{
{\bf The macroscopic reaction rate for the exterior of a ball.}  The
diffusive flux $J(t)$ onto the perfectly reactive boundary of a ball
of radius $R$ from Eq. (\ref{eq:Jt}), normalized by the steady-state
Smoluchowski rate $J_S$ from Eq. (\ref{eq:J_Smol}).  The disorder
scale $\mu$ takes 64 values in the logarithmic range from $10^{-1}$
(dark blue) to $10^1$ (dark red), as indicated in the colorbar.  The
shape parameter $\nu$ is set to $2$ {\bf (a)} or $0.5$ {\bf (b)}.  The
mean diffusivity $\bar{D}$ is fixed, whereas two other parameters of
the model are $\tau = R^2 \mu^2 \nu/\bar{D}$ and $\sigma = \bar{D}/(R
\mu \nu)$.  The upper black curve presents the flux $J_{\rm
hom}(t)/J_S$ from Eq. (\ref{eq:Jt_BM}) for homogeneous diffusion
(corresponding to $\mu = 0$), whereas the lower red curve shows the
asymptotic relation (\ref{eq:Jt_short}) corresponding to the limit
$\mu\to\infty$.}
\label{fig:Jt}
\end{figure}

Using the relation (\ref{eq:qT_tilde}), one can rewrite the integral
in Eq. (\ref{eq:Jt}) as
\begin{equation}  \label{eq:auxil12}
\int\limits_0^\infty dq \, \frac{\Upsilon'(t;q^2)}{q^2} = - \frac{\sqrt{\pi}}{2} \int\limits_0^\infty dT \, \frac{q(t;T)}{\sqrt{T}} \,,
\end{equation}
where $q(t;T)$ is the probability density function of the first moment
$t$ when the integrated diffusivity $T_t$ crosses the level $T$ (see
Sec. \SIsubordination).  In the short-time limit, one can resort again
to the superstatistical approximation by setting $T_t \simeq D t$,
with $D$ randomly drawn from the Gamma distribution (\ref{eq:Gamma}).
In other words, we approximate $q(t;T)$ as $q(t;T) \approx \langle
\delta(t-T/D) \rangle$, where the average is over all random
realizations of $D$.  This gives the following short-time
approximation:
\begin{equation}
q(t;T) \approx \frac{1}{\Gamma(\nu) t}\left(\frac{\nu T}{\bar{D}t}\right)^{\nu} \exp\left(-\frac{\nu T}{\bar{D}t}\right)\;.
\end{equation}
Substitution of this approximation into Eq. (\ref{eq:auxil12}) results
in the short-time asymptotic behavior of the rate:
\begin{equation}   \label{eq:Jt_short}
J(t) \simeq 4\pi c_0 R \bar{D} \biggl(1 + \frac{\Gamma(\nu+1/2)}{\sqrt{\nu} \, \Gamma(\nu)} \, \frac{R}{\sqrt{\pi \bar{D} t}}\biggr) 
\qquad (t\to 0).
\end{equation}
This relation is close to Eq. (\ref{eq:Jt_BM}) for Brownian motion
with mean diffusivity $\bar{D}$, in which the divergent $t^{-1/2}$
term is multiplied by the explicit prefactor
$\frac{\Gamma(\nu+1/2)}{\sqrt{\nu} \, \Gamma(\nu)}$ depending only on
$\nu$.  This prefactor monotonously grows from $0$ to $1$ as $\nu$
increases from $0$ to infinity (the limit $\nu \to \infty$
corresponding to Brownian motion).  As a consequence, the dynamic
disorder tends to diminish the macroscopic reaction rate, in agreement
with our statement about an increase of the mean FPT in bounded
domains.  In turn, the impact of disorder for unbounded domains is
rather weak, for instance, the prefactor is $0.8$ for $\nu = 1/2$.
The approximate asymptotic relation (\ref{eq:Jt_short}) does not
depend on the disorder scale $\mu$.  According to the superstatistical
approximation, this relation is actually the lower bound for the flux
$J(t)$ corresponding to the limit $\tau\to 0$ or, equivalently,
$\mu\to\infty$.  In turn, the exact expression (\ref{eq:Jt_BM}) for
Brownian motion corresponds to the limit $\mu\to 0$ and thus is close
to the upper bound for $J(t)$.  Although the flux is not necessarily a
monotonous function of $\mu$, this qualitative analysis accurately
describes the behavior of the flux $J(t)$.
Figure \ref{fig:Jt} shows the macroscopic reaction rate $J(t)$ from
Eq. (\ref{eq:Jt}) normalized by its steady-state value $J_S$.  For
both weak ($1/\nu = 0.5$) and strong ($1/\nu = 2$) disorder, a
substantial increase of the disorder scale $\mu = \sigma\tau/R$ from
$10^{-1}$ to $10^{1}$ has only a minor effect, and all curves are
close to both the classic flux $J_{\rm hom}(t)$ from
Eq. (\ref{eq:Jt_BM}) and the asymptotic relation (\ref{eq:Jt_short}).

We stress that the total flux $J(t)$ was computed by integrating
the probability fluxes $\rho(t|\x_0)$.  In turn, the common way of
obtaining the total flux consists in finding the concentration profile
$c(\x,t)$ and then integrating the diffusive flux density, $-D
\partial c(\x,t)/\partial n$, over the target surface $\pa$, where
$\partial/\partial n$ is the normal derivative oriented outward the
confining domain.  However, the diffusivity $D$ is random in the
annealed model of heterogeneous diffusion that prohibits using the
above form of the diffusive flux density.  If the random $D$ is
replaced by the mean diffusivity $\bar{D}$, the total flux could then
be approximated as
\begin{eqnarray}  \nonumber
J_{\rm app}(t) &=& \int\limits_{\pa} d\s \biggl. 
\biggl(- \bar{D} \frac{\partial c_0 S(t|\x_0)}{\partial n}\biggr)\biggr|_{\x_0 = \s} \\   \label{eq:Jt_approx}
&=& 4\pi R c_0 \bar{D} \biggl(1 + \frac{2R}{\pi} \int\limits_0^\infty dq \, \Upsilon(t;q^2) \biggr).
\end{eqnarray}
While the long-time limit of the flux (the first term) is the same as
in the exact solution (\ref{eq:Jt}), the approach to this limit, given
by the second term, is different.  The formulas (\ref{eq:Jt},
\ref{eq:Jt_approx}) are identical only for homogeneous diffusion.
This computation illustrates some pitfalls of applying conventional
tools of homogeneous diffusion to heterogeneous one.

\begin{figure}[t!]
\begin{center}  
\includegraphics[width=85mm]{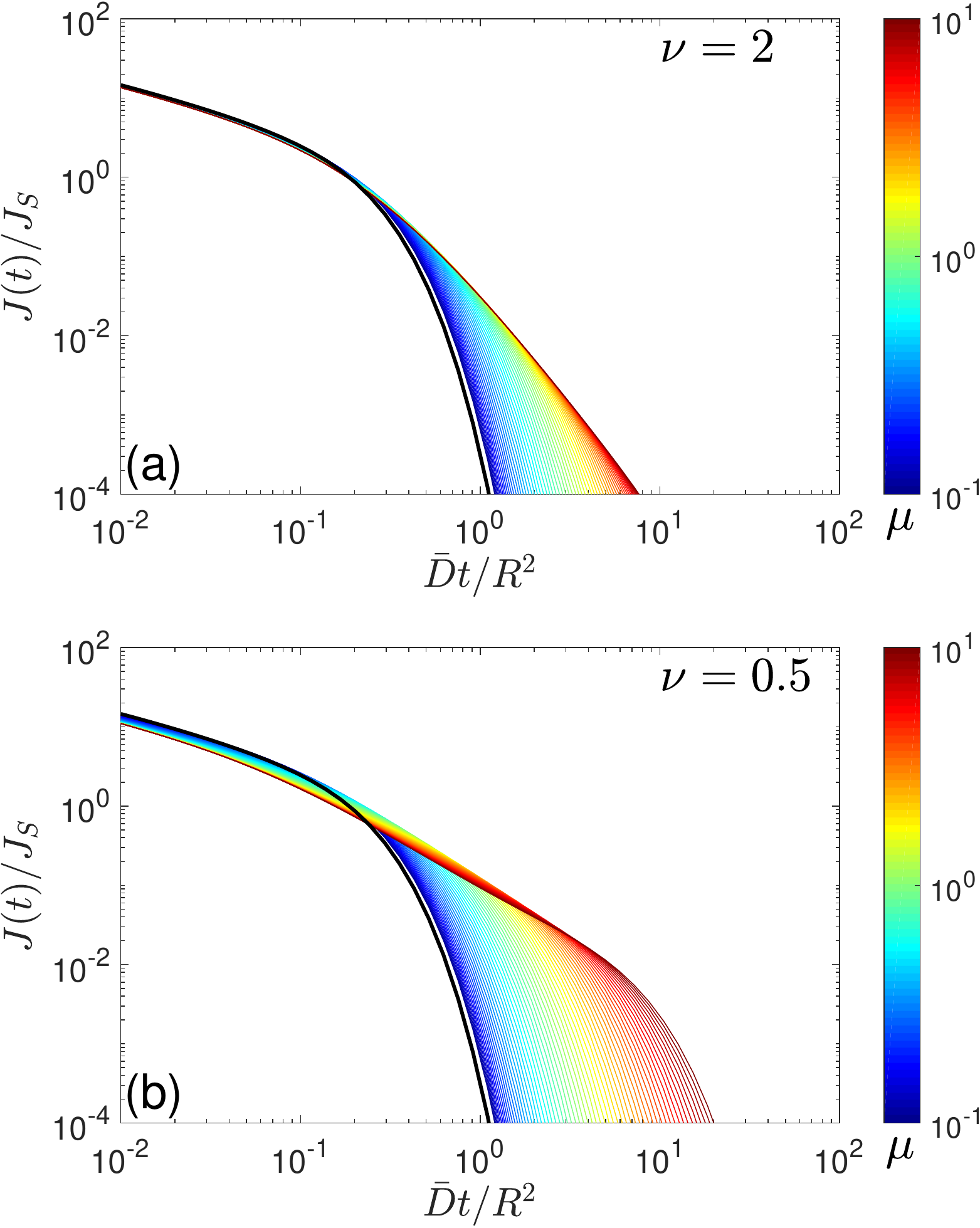}
\end{center}
\caption{
{\bf The macroscopic reaction rate for the interior of a ball.}  The
diffusive flux $J(t)$ onto the perfectly reactive surface of a ball of
radius $R$, from Eq. (\ref{eq:Jsphere}), normalized by the
Smoluckowski steady-state rate $J_S$ from Eq. (\ref{eq:J_Smol}), with
the uniform initial concentration $c_0$ inside the ball, $\nu = 2$
{\bf (a)} and $\nu = 0.5$ {\bf (b)} and 64 values of $\mu$ in the
logarithmic range between $10^{-1}$ and $10^1$ (curves encoded by
color, ranging from dark blue to dark red, as shown by colorbar).
Thick black line shows the macroscopic reaction rate for homogeneous
diffusion with diffusivity $\bar{D}$. }
\label{fig:Jsphere}
\end{figure}

For comparison, we also compute the macroscopic reaction rate for
heterogeneous diffusion inside a ball of radius $R$.  From
Eq. (\ref{eq:J_main}), one gets
\begin{equation}  \label{eq:Jsphere}
J(t) = - \frac{8c_0 R^3}{\pi} \sum\limits_{n=1}^\infty \frac{\Upsilon'(t; \pi^2 n^2/R^2)}{n^2} \,,
\end{equation}
where $c_0$ is the uniform initial concentration.  Figure
\ref{fig:Jsphere} shows the behavior of this rate, normalized for
convenience by the Smoluchowski steady-state rate $J_S$ from
Eq. (\ref{eq:J_Smol}).  As this confining domain is bounded, the
reaction rate vanishes at long times, as all particles will finally
react.  This is an evident difference from Fig. \ref{fig:Jt}, in which
the reaction rate reaches a nonzero limit $J_S$.  For a fixed disorder
strength $1/\nu$, the curves exhibit a much stronger dependence on the
disorder scale $\mu$ for interior diffusion than for exterior one.
This observation re-confirms that the dynamic disorder is averaged
more efficiently in unbounded domains.  In turn, one observes in both
Fig. \ref{fig:Jt} and Fig. \ref{fig:Jsphere} a broader dispersion of
curves for stronger disorder.  Finally, one sees that the dynamic
disorder leads to a higher reaction rate at long times, in agreement
with our conclusion that the reaction kinetics is slowed down on
average and thus more particles remain present in the confining
domain.


\paragraph*{Collective search by multiple independent particles.}
 
The description of a single particle opens a way to investigate some
basic multi-particle effects.  For instance, when $N$ independent
particles simultaneously search for a target, the distribution of the
first arrival is still determined by the survival probability for a
single particle \cite{Yuste2007,Grebenkov2010b}.  If the starting
points of these particles are uniformly distributed in a region
$\Omega = \{ \x\in \R^3~:~ R < \|\x\| < R_{\rm max}\}$ around the
spherical target of radius $R$, one gets
\begin{equation}
S_N(t) = \P\{ \min\{\T_1,\ldots,\T_N\} > t\} = \biggl( \int\limits_\Omega \frac{d\x_0}{V} \, S(t|\x_0)\biggr)^N ,
\end{equation}
where $V$ is the volume of $\Omega$, and $\T_1,\ldots,\T_N$ are
independent first-passage times for $N$ particles.  In the
thermodynamic limit, when both $N$ and $V$ (or $R_{\rm max}$) tend to
infinity but the density $c_0 = N/V$ remains fixed, one finds
\begin{equation}
-\ln(S_\infty(t)) = c_0 \int\limits_{\|\x_0\|>R} d\x_0 \, \bigl(1 - S(t|\x_0)\bigr).
\end{equation}
The right-hand side is the number of particles that reacted up to time
$t$ which can also be obtained by integrating the flux $J(t)$ from
Eq. (\ref{eq:Jt}):
\begin{eqnarray}  
&& S_\infty(t) =  \exp\biggl(- \int\limits_0^t dt' \, J(t')\biggr) \\  \nonumber
&& = \exp\biggl( - 4\pi c_0 R \bar{D} \biggl(t + \frac{2R}{\pi \bar{D}} \int\limits_0^\infty dq \, \frac{1 - \Upsilon(t;q^2)}{q^2} \biggr)\biggr)  .
\end{eqnarray}
Once again, this quantity is fully determined by
$\Upsilon(t;\lambda)$.  In particular, one can use the above
asymptotic relations to study the behavior of $S_\infty(t)$ at short
and long times.

We note, however, that the validity of the assumption of independent
particles is debatable in the context of dynamically re-arranging
media.  In fact, when two particles come close to each other, they
probe the same local environment and thus should have similar
diffusivities.  As a consequence, the stochastic diffusivities of
these particles become correlated (locally in time).  The impact of
this intricate correlation mechanism onto diffusion-controlled
reactions remains an open challenging problem for future
investigations.  We also note that the same issue concerns the
macroscopic reaction rate $J(t)$ in Eq. (\ref{eq:J_main}), which is
obtained by superimposing probability fluxes $\rho(t|\x_0)$ from
independent particles with a prescribed initial concentration
$c_0(\x_0)$.

\subsection*{5. Subordination approach}

The subordination approach consists in treating time-dependent
diffusivity as changing the ``internal time'' of the process
\cite{Chechkin2017}.  When the diffusivity $D_t$ is deterministic, the
diffusion equation for the propagator, $\partial P/\partial t = D_t
\Delta P$, can be reduced to the ``canonical'' form, $\partial P_{\rm
hom}/\partial T_t = \Delta P_{\rm hom}$ with unit diffusivity, where a
new ``internal time'' variable is
\begin{equation}
T_t = \int\limits_0^t dt' \, D_{t'} 
\end{equation}
(we call $T_t$ ``internal time'', in spite of its units m$^2$; to get
usual time units, one can divide it by $\bar{D}$ or another
diffusivity).  In other words, time-dependent diffusion can be
understood as traveling along a random path generated by ordinary
Brownian motion, but with a variable, time-dependent ``speed''
$dT_t/dt = D_t$.

The same argument holds for stochastic diffusivity $D_t$, in which
case the internal time $T_t$ is a stochastic process.  The
conventional spectral decomposition of the propagator in the internal
time $T_t$,
\begin{equation}  \label{eq:Phom_auxil}
P_{\rm hom}(\x, T_t | \x_0) = \sum\limits_{n=1}^\infty u_n(\x) \, u_n(\x_0) \, e^{- \lambda_n T_t} \,,
\end{equation}
should be averaged with the probability density function $Q(t;T)$ of
the integrated diffusivity $T_t$:
\begin{eqnarray}   \label{eq:P_sub}
P(\x, t | \x_0) &=& \langle P_{\rm hom}(\x,T_t|\x_0)\rangle_{T_t} \\  \nonumber
&=& \int\limits_0^\infty dT \, Q(t;T) \, P_{\rm hom}(\x, T|\x_0) \\   \nonumber
&=& \sum\limits_{n=1}^\infty u_n(\x) \, u_n(\x_0) \, \underbrace{\int\limits_0^\infty dT \, Q(t;T) e^{-\lambda_n T}}_{\Upsilon(t; \lambda_n)} \,.
\end{eqnarray}
One gets therefore the natural interpretation (\ref{eq:QT_def}) of
$\Upsilon(t;\lambda)$ as the Laplace transform of the probability
density function $Q(t;T)$ of the integrated diffusivity $T_t$.  The
related first-passage times for the Feller process were investigated
\cite{Masoliver2012}.  The probability density function $Q(t;T)$
``translates'' the internal time $T$ into the physical time $t$.  If
$D_t$ is deterministic, then $Q(t;T) = \delta(T - T_t)$ and thus
$\Upsilon(t;\lambda) = \exp(-\lambda T_t)$, as expected.  Note that
here we considered the internal time $T_t$ averaged over the initial
diffusivity $D_0$ drawn from the stationary Gamma distribution
(\ref{eq:Gamma}).  In turn, if $D_0$ is fixed, the function
$\Upsilon(t;\lambda)$ is replaced by $\Upsilon(t;\lambda| D_0)$ from
Eq. (\ref{eq:Upsilon_D0}), which is the Laplace transform of the
corresponding probability density function $Q(t;T|D_0)$.

The numerical computation of the probability density function $Q(t;T)$
would require the inversion of the Laplace transform.  In turn, the
moments of the integrated diffusivity $T_t$ can be easily obtained via
Eq. (\ref{eq:QT_def}):
\begin{equation}
\langle (T_t)^k \rangle = (-1)^k \lim\limits_{\lambda\to 0} \frac{\partial^k \Upsilon(t;\lambda)}{\partial \lambda^k} \,.
\end{equation}
In particular, the mean and the variance are
\begin{equation}
\langle T_t \rangle = \bar{D} t , \qquad \var\{T_t\} = \frac{2\tau^2 \bar{D}^2}{\nu} \bigl(t/\tau - 1 + e^{-t/\tau}\bigr).
\end{equation}
As expected, the mean integrated diffusivities for heterogeneous and
homogeneous diffusions are identical and grow linearly with time.  The
variance grows quadratically at small times ($t \ll \tau$) and
linearly at large times ($t \gg \tau$).  As a consequence, the squared
coefficient of variation, 
\begin{equation}  \label{eq:varT}
\frac{\var\{T_t\}}{\langle T_t \rangle^2} = \frac{2\tau}{\nu t} \biggl(1 - \frac{\tau}{t} \bigl(1 - e^{-t/\tau}\bigr)\biggr),
\end{equation}
monotonously decreases from $1/\nu$ at $t= 0$ to zero as $t\to\infty$.
The shape parameter $\nu$ thus controls the broadness of the
distribution of $T_t$ at short times, given that the initial
diffusivity $D_0$ is randomly picked up from the Gamma distribution
(\ref{eq:Gamma}), see Eq. (\ref{eq:one_nu}).  Note also that the
right-hand side of Eq. (\ref{eq:varT}) coincides with the non-Gaussian
parameter for the one-dimensional heterogeneous diffusion in the free
space $\R$ \cite{Lanoiselee2018}.

In the same vein, the first-passage time $\T$ to a reactive target can
be related to the first-crossing time of a random barrier by the
stochastic process $T_t$ (Fig. \ref{fig:subord}).  In fact, one can
first generate a random path to the target by ordinary Brownian motion
with unit diffusivity and then consider a particle traveling along
this path with a time-dependent ``speed''.  The target is reached when
the whole path is passed through, i.e., when the internal time $T_t$
attains the duration $\T_{\rm hom}$ of the Brownian path, which is
random and determined by the conventional probability density function
of the first-passage time for Brownian motion with unit diffusivity
\begin{align}  
\rho_{\rm hom}(T|\x_0) &= \frac{\P_{\x_0}\{ \T_{\rm hom} \in (T,T+dT)\}}{dT} \nonumber   \\   \label{eq:rho_hom}
&= \sum\limits_{n=1}^\infty u_n(\x_0) \lambda_n e^{-\lambda_n T} \int\limits_\Omega d\x \, u_n(\x).  
\end{align}
Let $\delta_T = \inf\{ t > 0~:~ T_t > T\}$ to be the random time when
the process $T_t$ crosses a fixed level $T$.  Since $T_t$ monotonously
increases, one has
\begin{equation}
\P\{ \delta_T > t\} = \P\{ T_t < T\} = \int\limits_0^T dT' \, Q(t;T').
\end{equation}
In particular, the probability density function of the random time
$\delta_T$ reads
\begin{equation}  \label{eq:qT_def}
q(t;T) = - \frac{\partial  \P\{ \delta_T > t\}}{\partial t} = - \int\limits_0^T dT' \, \frac{\partial Q(t;T')}{\partial t} \,.
\end{equation}
If now the level $T$ is the random duration of the Brownian path, $T =
\T_{\rm hom}$, the random time $\T = \delta_{\T_{\rm hom}}$ is the
first-passage time to the reactive target, and its probability density
function is obtained by averaging $q(t;\T_{\rm hom})$ over the
distribution of $\T_{\rm hom}$:
\begin{equation}  \label{eq:rho_sub}
\rho(t|\x_0) = \int\limits_0^\infty dT \, q(t;T) \, \rho_{\rm hom}(T|\x_0) .
\end{equation}
Substitution of Eq. (\ref{eq:rho_hom}) into this relation allows one
to retrieve Eq. (\ref{eq:rho_main}).  Multiplying
Eq. (\ref{eq:qT_def}) by $e^{-\lambda T}$ and integrating over $T$
from $0$ to infinity, one can express the Laplace transform of the
density $q(t;T)$ as
\begin{equation}  \label{eq:qT_tilde}
\int\limits_0^\infty dT \, e^{-\lambda T} \, q(t;T) = - \frac{\Upsilon'(t;\lambda)}{\lambda} \,,
\end{equation}
where Eq. (\ref{eq:QT_def}) was used.  One can see that the function
$\Upsilon(t;\lambda)$ and its time derivative $\Upsilon'(t;\lambda)$,
explicitly known from Eqs. (\ref{eq:Ups},
\ref{eq:dP}), fully determine the densities $Q(t;T)$ and $q(t;T)$ via
Laplace transforms.


In summary, Eqs. (\ref{eq:P_sub}, \ref{eq:rho_sub}) couple the spatial
aspects of the problem (such as the geometric structure of the medium,
the shape, location and reactivity of the targets, and the starting
point) to the dynamic disorder represented by the stochastic
diffusivity.  The spatial features do not depend on diffusivity and
are determined by homogeneous diffusion (ordinary Brownian motion).
In turn, the disorder aspects are captured via the distribution of the
integrated diffusivity $T_t$.  Although we focused on the stochastic
diffusivity modeled by the Feller process (\ref{eq:Feller}), one can
explore other models such as, e.g., reflected Brownian motion
\cite{Chubynsky2014}, L\'evy-driven stochastic diffusivity
\cite{Jain2017a} or geometric Brownian motion.  

On the other hand, the subordination approach is limited to the
marginal propagator $P(\x,t|\x_0)$ and related quantities (such as the
probability density function $\rho(t|\x_0)$) and does not yield the
full propagator $P(\x,D,t|\x_0,D_0)$ that we obtained in
Sec. \SIgeneral by solving the Fokker-Planck equation.  We also stress
that the subordination does not resolve the problem of partially
reactive targets with Robin boundary condition, as discussed in the
Discussion section.  In fact, even though the conventional spectral
decomposition (\ref{eq:Phom_auxil}) is valid for partially reactive
boundary condition, its formal extension to heterogeneous diffusion
via the subordination (\ref{eq:P_sub}) remains debatable as it would
correspond to a modified model of stochastic diffusivity $D_t$, in
which $D_t$ should take a fixed prescribed value when the particle is
on the target.  A proper description of heterogeneous diffusion toward
partially reactive targets remains an open mathematical problem.

\section*{Data Availability}

Data sharing not applicable to this article as no datasets were
generated or analysed during the current study.

\section*{Code Availability}

All figures have been prepared by means of Matlab software.  The
plotted quantities have been computed by explicit formulas provided in
the letter by using custom routines for Matlab software.  While the
explicit form makes these numerical computations straightforward,
custom routines are available from the corresponding author upon
request.

\section*{END NOTES}

\begin{acknowledgments}
D.S.G. acknowledges the support under Grant No. ANR-13-JSV5-0006-01 of
the French National Research Agency.
\end{acknowledgments}

\section*{Author Contributions} 
Y.L. and D.S.G. designed research; Y.L., N.M., and D.S.G. performed
research and analyzed the results; D.S.G. wrote the paper.

\section*{Competing Interests Statement}
The authors declare no competing interests.

\section*{Corresponding Author}
The corresponding author is Denis S. Grebenkov
(denis.grebenkov@polytechnique.edu).

\newpage 

\newpage

\newpage


\newpage

\newpage

\newpage

\newpage

\newpage

\end{document}